\documentclass[epjST]{svjour}
\usepackage{amsmath}
\usepackage{amssymb}
\usepackage{graphicx}
\usepackage{cite}
\usepackage{braket}
\usepackage{bm}
\usepackage{hyperref}
\DeclareMathOperator{\sign}{sgn}
\newcommand{\abs}[1]{\left| #1 \right|}

\begin{document}
\title{Transport signatures of Majorana bound states in superconducting hybrid structures}
\subtitle{A minireview}
\author{Alexander Schuray\inst{1} \and Daniel Frombach\inst{1} \and Sunghun Park\inst{2} \and Patrik Recher\inst{1,3} }
\institute{Institute for Mathematical Physics, TU Braunschweig, D-38106  Braunschweig, Germany \and Departamento de F\'isica Te\'orica de la Materia Condensada, Condensed Matter Physics Center (IFIMAC) and Instituto Nicol\'as Cabrera, Universidad Aut\'onoma de Madrid, Spain \and Laboratory for Emerging Nanometrology Braunschweig, D-38106 Braunschweig, Germany}
\abstract{In this minireview, we outline the recent experimental and theoretical progress in the creation, characterization and manipulation of Majorana bound states (MBSs) in semiconductor-superconductor (SC) hybrid structures. After an introductory overview of the broader field we specifically focus on four of our recent projects in this direction. We show that the emergence of Fano resonances in the differential conductance in a normal lead-Majorana nanowire-quantum dot setup can be exploited to determine if a single MBS is contacted by the normal lead and the quantum dot providing an experimental test of the non-locality of MBSs. In the second project, the tunnel-coupling to two MBSs in an $s$-wave SC-Majorana nanowire Josephson junction (JJ) leads to a finite contribution of the MBSs to the equilibrium Josephson current probing directly the local spin-singlet contribution of the Majorana pair. We then shift our focus from MBSs forming in nanowire systems to MBSs forming in topological JJs.
In a single sheet of buckled silicene with proximity induced superconductivity two local electric fields can be used to tune the junction between a topologically trivial and topologically non-trivial regime. In a Corbino geometry topological Josephson junction two MBSs harbored in Josephson vortices can rotate along the JJ and, in the course of this, will be exchanged periodically in the phase difference of the JJ. The tunneling current in a metal tip coupled to the JJ is shown to exhibit signs of the anyonic braiding phase of two MBSs.
} %
\maketitle
\section{Introduction}
\label{intro}
In this minireview, we give an overview of the current investigations towards the creation, detection and manipulation of Majorana bound states (MBSs)
appearing in superconducting hybrid junctions. We then discuss four projects where we propose specific ways to characterize and manipulate MBSs in transport setups contributing towards the goal of using such non-Abelian zero modes for qubits in topological quantum computation (TQC). The potential interest of such MBSs lies in the ability to store a single Dirac fermionic degree of freedom (DFDF) with annihilation operator $c$ in two MBSs $\gamma_1$ and $\gamma_2$ via the relation $c=(\gamma_1+i\gamma_2)/2$ where $\gamma_{1,2}^{\dagger}=\gamma_{1,2}$ are hermitian~\cite{Kitaev2001}. They can be spatially distant such that no local measurement can access the quantum numbers associated with a single DFDF which is its occupation $c^{\dagger}c$ having eigenvalues 0 or 1 and that can be associated with a unit of information (like in a qubit). The formal process to decompose the DFDF into two MBSs is called fusion and they can be considered (loosely speaking) as the real and imaginary part of that DFDF. To make topologically protected operations, the MBSs can be pairwise braided~\cite{Kitaev2003,Nayak2008,Pachos2012,Lahtinen2017}. The braiding statistics is that of Ising anyons which is neither bosonic (relativ phase $0$) nor fermionic (relativ phase $\pi$) but has a relative braiding phase of $\pi/2$~\cite{Ivanov2001,Nayak2008,Alicea2011}. \footnote{The exchange of two MBSs {\it twice} introduces a minus sign in both MBSs ($\gamma_1 \rightarrow -\gamma_1$, $\gamma_2 \rightarrow -\gamma_2$). In that sense, the single exchange can be associated with a phase of $\pi/2$} Such particles which are neither fermions nor bosons are called anyons~\cite{Wilczek2009}. It is the non-Abelian nature of the Ising anyons that allows for quantum state changes by braiding of MBSs in a degenerate ground state manifold~\cite{Ivanov2001,Stern2004,Leijnse2012}, which we will explain in more details in Section~\ref{sec4:s-ti}.

MBSs have been predicted to exist in vortices of p-wave superconductors (SCs)~\cite{Read2000} which can be traced back to the winding in momentum space of the p-wave pairing amplitude. Since time-reversal symmetry is broken in such materials, the pairing potential is supposed to be vulnerable to disorder \cite{Beenakker2013b}. More recently, the seminal paper by Fu and Kane suggested a more practical way to produce an effective p-wave superconductor utilizing the proximity effect with an ordinary s-wave superconductor in contact with the surface of a 3D topological insulator~\cite{Fu2008} where the necessary winding is now provided by the Dirac surface electrons (acquiring a Berry phase of $\pi$ in a closed loop in real space). MBSs would then appear in vortices or at the boundaries with magnetic elements in such topological superconductors (TSCs).

Signatures of MBSs in such proximity structures were searched for early on theoretically and experimentally. Here, we only give a brief overview which also should set the stage for the projects that we will discuss in the later section of this minireview. Excellent and more extensive reviews exist in the literature where one can find further information and references~\cite{Hasan2010,Qi2011,Alicea2012,Leijnse2012,Beenakker2013b,Sato2017,Aguado2017,Lutchyn2018,Culcer2019}. 

The first systems that have been examined experimentally were the spin-orbit coupled quantum wires in proximity with an s-wave SC and subject to a magnetic Zeeman field~\cite{Mourik2012,Rokhinson2012,Deng2012,Das2012,Churchill2013,Finck2013,Lee2013,Albrecht2016,Deng2016,Chen2017,Suominen2017,Nichele2017,Guel2018,Zhang2018,Sestoft2018,Deng2018,Laroche2019}. In these wires, MBSs are predicted to appear at the wire's ends in the topological regime~\cite{Lutchyn2010,Oreg2010}. Other one-dimensional realizations contain the magnetic adatoms arranged in a chain on top of an $s$-wave superconductor~\cite{Choy2011,Nadj-Perge2013,Klinovaja2013,Braunecker2013,Vazifeh2013,Pientka2013,Nadj-Perge2014,Pawlak2016,Jeon2017,Ruby2017,Marra2019}. 

There are several ways to test the existence of such MBSs. The probably most straightforward way is to couple normal leads 
as a local probe. The predicted transport signature of a MBS at zero temperature is a zero-bias conductance resonance with value $2e^2/h$, independent on the tunneling strength~\cite{Bolech2007,Law2009,Flensberg2010,Prada2012,Stanescu2012,Rainis2013,Paul2018}. When the SC is floating, the charging energy starts to play a role at low temperatures and the signature of MBSs would be explicit in the Coulomb blockade peaks showing a period of $e$ and not $2e$ ($e$ the elementary charge) as a function of gate voltage~\cite{Fu2010,Zazunov2011,vanHeck2016}, which has been observed experimentally~\cite{Albrecht2016}. Coupling the MBSs to normal metal leads is certainly not the desired situation when the system should be topologically protected since electron tunneling changes the parity of the system and thereby can lead to dephasing~\cite{Budich2012}. However, the coupling to measuring leads can be used to inspect the Majorana system. In the nanowires, the MBSs appear at the ends of the wire, however, theoretical calculations show that the wave functions of the MBSs substantially leak into the bulk of the wire~\cite{DasSarma2012,Klinovaja2012}. When can we therefore consider two MBSs $\gamma_1$ and $\gamma_2$ as spatially separated such that the environment cannot couple to both MBSs simultaneously? In the desired case where parity is conserved, any perturbation coupling to the MBSs has to be a product of two MBSs (which is proportional to the parity operator ${\cal P}=i\gamma_1\gamma_2$ for the occupation of the DFDF defined by the two MBSs). Any (parity-conserving) perturbation therefore would have to couple to both MBSs and should be exponentially suppressed with the distance of the two MBSs for perturbations acting locally~\cite{Kitaev2001,Leijnse2012}. If local perturbations can couple to both MBSs is therefore an indispensable insight for practical purposes. A way to test this feature is to look at (Fano)-resonances in a setup where a Majorana nanowire is in addition coupled to a quantum dot~\cite{Schuray2017} as we will explain in more details in Section~\ref{Sec:Fano}. Although MBSs are spinless particles, their electron-like (or hole-like) spinor components have finite spin expectation values that depend on various system parameters~\cite{Sticlet2012,Serina2018}. Especially, the spinors for MBSs in the nanowires have been analyzed in detail and proposals for the detection of their form include scanning probe signals~\cite{Chevallier2016} and magnetic field dependent spectroscopy using a quantum dot~\cite{Prada2017,Penaranda2018}. In Section~\ref{Sec:spincanting}, we will discuss additional spin-properties in an s-wave SC-Majorana nanowire Josephson junction (JJ) where the Josephson effect tests the spatial overlap of {\it two} MBSs in the singlet channel~\cite{Schuray2018}.

Other decisive signatures are provided by the fractional Josephson effect, which is predicted to exhibit a $4\pi$-periodic phase dependence when two MBSs from different SCs are coupled by electron tunneling~\cite{Kitaev2001,Kwon2004,Fu2009b,Alicea2011,Kane2015, Crepin2014}. The doubled period can be explained by noting that the two MBSs connecting the two TSCs allow for transfer of single electrons rather than Cooper pairs such that the phase difference of the JJ is effectively only half the usual value so that the current phase relation assumes the form $I=I_c\sin(\phi/2)$ where $I_c$ is the critical current and $\phi$ is the SC phase difference across the JJ. This seems like an easy-to-verify relation and a very characteristic one as well. However, in reality the number parity which protects this result is challenged by quasiparticle poisoning~\cite{Pikulin2012,Rainis2012}. In addition, the finite size of the TSCs introduces two additional MBSs which leads to a gap opening at the parity crossing~\cite{San-Jose2012,Houzet2013,Virtanen2013,Cayao2017,Cayao2018b}. Such fractional phase dependences have been studied in the frequency domain (Shapiro effect ~\cite{Rokhinson2012,Bocquillon2016,Wiedenmann2016,Bocquillon2018,Dominguez2012,Dominguez2017,Pico-Cortes2017} and Josephson emission/radiation~\cite{Deacon2017,Laroche2019}) where the Josephson junction is brought out of equilibrium. Distinguishing signatures of JJs in the topological regime were also predicted via dc-measurements in the long junction regime (the length of the JJ is larger than the superconducting coherence length)~\cite{Beenakker2013}. In the latter case, it was shown theoretically that the critical current of a JJ (an equilibrium property) in the long junction regime is twice as large if it behaves as a $4\pi$-junction compared to the more common $2\pi$-junction. To compare the two regimes, one would need to have a JJ that can be tuned between the two cases~\cite{Frombach2018}. This we will discuss in Section~\ref{Sec:silicene}. We also note that Josephson junctions in two dimensional (2D) electron gases with spin-orbit coupling can lead to the emergence of MBSs~\cite{Mi2013,Kuzmanovski2016,Li2016b,Pientka2017,Hell2017,Finocciaro2018,Ren2019,Scharf2019}.

The final goal, however, will be to show the non-Abelian statistics of MBSs since this feature is probably the most exotic and also the most useful one thinking in terms of using such states for TQC~\cite{Ivanov2001,Sarma2015}. There are several proposals how to braid Majorana fermions~\cite{Alicea2011,Hoffman2016,Karzig2017,Plugge2017} in superconducting hybrid systems. In the network of wires, this is not trivial since the nanowires are one-dimensional and one has to think about effectively 2D systems like a T-junction \cite{Alicea2011,Sau2011,Aasen2016,Litinski2017}. In this sense, the 2D versions of TSCs seem more feasible. Here, MBSs can be isolated in vortices, however it is not straightforward to think about ways to move these vortices. In Section~\ref{sec4:Corbino}, we will review our proposal to move two MBSs in a Corbino geometry JJ and to detect their exchange phase via the current in a STM tip. We note that braiding alone cannot provide universal quantum computation with MBSs as the rotation angles are fixed and so not all single-particle rotations can be covered by braiding~\cite{Bravyi2005,Bravyi2006}. Also, entanglement between qubits is inaccessible by braiding alone.For more information on TQC, see the more extensive review on this topic~\cite{Nayak2008,Pachos2012}. We note that braiding operations of MBSs can also be performed without physically moving the MBSs in real space, but by a sequence of switching on and off the couplings between Majoranas~\cite{Hassler2011,vanHeck2012} or by making a sequence of projection measurements~\cite{Bonderson2008}. A recent review of the possible ways to braid and fuse MBSs in superconducting hybrid junctions is given in~\cite{Beenakker2019}. Generalized Majorana fermions, so called parafermions, are more exhaustive in terms of possible braiding operations as compared to Majorana fermions and could serve as building blocks for universal TQC \cite{Nayak2008, Hutter2016}. Realistic proposals for the creation of parafermions combine superconductivity with repulsive electron-electron interactions (for a review see Ref.~\cite{Alicea2016}). So far these fascinating but also challenging ideas await experimental realizations.

We would like to end this introduction by elaborating briefly on the challenges that remain in the identification of MBSs and what the experimental strategies are to deal with them. Although some of the key predictions like tunneling-conductance quantization~\cite{Zhang2018} or the observation of missing odd Shapiro steps~\cite{Rokhinson2012,Bocquillon2016,Wiedenmann2016,Bocquillon2018} have been seen in experiments such features were predicted to occur also in non-topological hybrid junctions. One of the main issues is the distinction of true {\it isolated} MBSs from the conventional Andreev bound states that are Dirac fermions with Majorana components that are not efficiently delocalized 
(e.g. by the length of a nanowire) and can appear in trivial systems~\cite{Cayao2015,SanJose2016,Liu2017,Liu2018,Vuik2018a,Avila2019,Reeg2018a, Moore2018a,Moore2018b,Chiu2019,Awoga2019,Woods2019a,Chen2019}. They are sometimes dubbed quasi-MBSs. Despite this difference, they can mimic the essential features of true MBSs including conductance quantization, the fractional Josephson effect and even braiding \cite{SanJose2016, Vuik2018a, Chiu2019}. This is possible e.g. by the scheme one uses to measure the MBSs features --- coupling measuring leads to the device results in an open quantum system. This can lead to very different couplings to the two components of Andreev bound states or quasi-MBSs and the measurement would appear as that of probing a TSC~\cite{Avila2019}. Experimental strategies to discriminate quasi-MBSs from true MBSs is to perform non-local measurements~\cite{Bolech2007,Nilsson2008,Moore2018a} which seems one of the next steps that should be feasible to do in the near future~\cite{Zhang2019}.
\section{Fano resonances in hybrid Majorana systems}
\label{Sec:Fano}
One of the first suggested signatures of an isolated MBS in one-dimensional TSCs~\cite{Kitaev2001,Lutchyn2010,Oreg2010,Nadj-Perge2013} is the quantization of the differential conductance when contacting the MBS with a metallic lead~\cite{Bolech2007,Law2009,Flensberg2010,Prada2012,Stanescu2012,Rainis2013}. And indeed several early experiments reported a zero bias peak in the differential conductance~\cite{Mourik2012,Das2012,Lee2013,Nadj-Perge2014,Pawlak2016}, however the quantization of this peak has been measured only recently~\cite{Zhang2018}. This absence of an unambiguous signature leads to the desire of extended experimental designs with superior tunability. Setups where quantum dots are tunnel-coupled to superconductors have been investigated as versatile devices in the past using conventional superconductors~\cite{Rodero2011}. Such quantum dot-superconductor hybrid systems can now also be considered in the case of topological superconductors, where MBSs hybridize with a quantum dot in the normal state~\cite{Liu2011,Leijnse2011,Vernek2014,Prada2017,Clarke2017,Ptok2017,Chevallier2018,Fleckenstein2018}.
In addition, the coupling of a quantum dot can lead to the emergence of Fano resonances in the differential conductance~\cite{Dessotti2014,Ueda2014,Xia2015,Jiang2015,Zeng2016,Xiong2016,Baranski2016,Wang2018b,Ricco2018}. First experiments using semiconductor-superconductor hybrid structures have already suggested the coupling of a quantum dot with MBSs~\cite{Deng2016,Deng2018}. \\
The differential conductance measured in these experiments revealed a hybridization between the near zero energy modes and the electron- and hole-like states on the quantum dot, but these hybridizations were different which does not fit with hybridization with a single MBS. However, it can be explained considering a coupling to the second more distant MBS~\cite{Prada2017,Schuray2017,Clarke2017,Ricco2019}. Various quality factors which measure the ``Majorananess'' of the coupling to the low-energy states have been proposed~\cite{Prada2017,Clarke2017,Penaranda2018}. \\
In this section, we want to report on our findings already investigated in reference~\cite{Schuray2017}, in which we considered a setup where two MBSs are contacted by a normal conducting lead on one side and a quantum dot on the other side. By including not only a coupling to the closest MBS we can tune the system between Majorana-like (only coupling to one MBS) and Dirac-like (coupling to both MBSs with the same amplitude) couplings. We show that pairs of Fano resonances arise as a function of the quantum dot level energy and that in the case of pure Majorana-like coupling these pairs are symmetric with respect to each other which reflects the electron-hole symmetry of the MBS.

\subsection{Model and cumulant generating function}
\begin{figure}
	\centering
	\includegraphics[width=\textwidth]{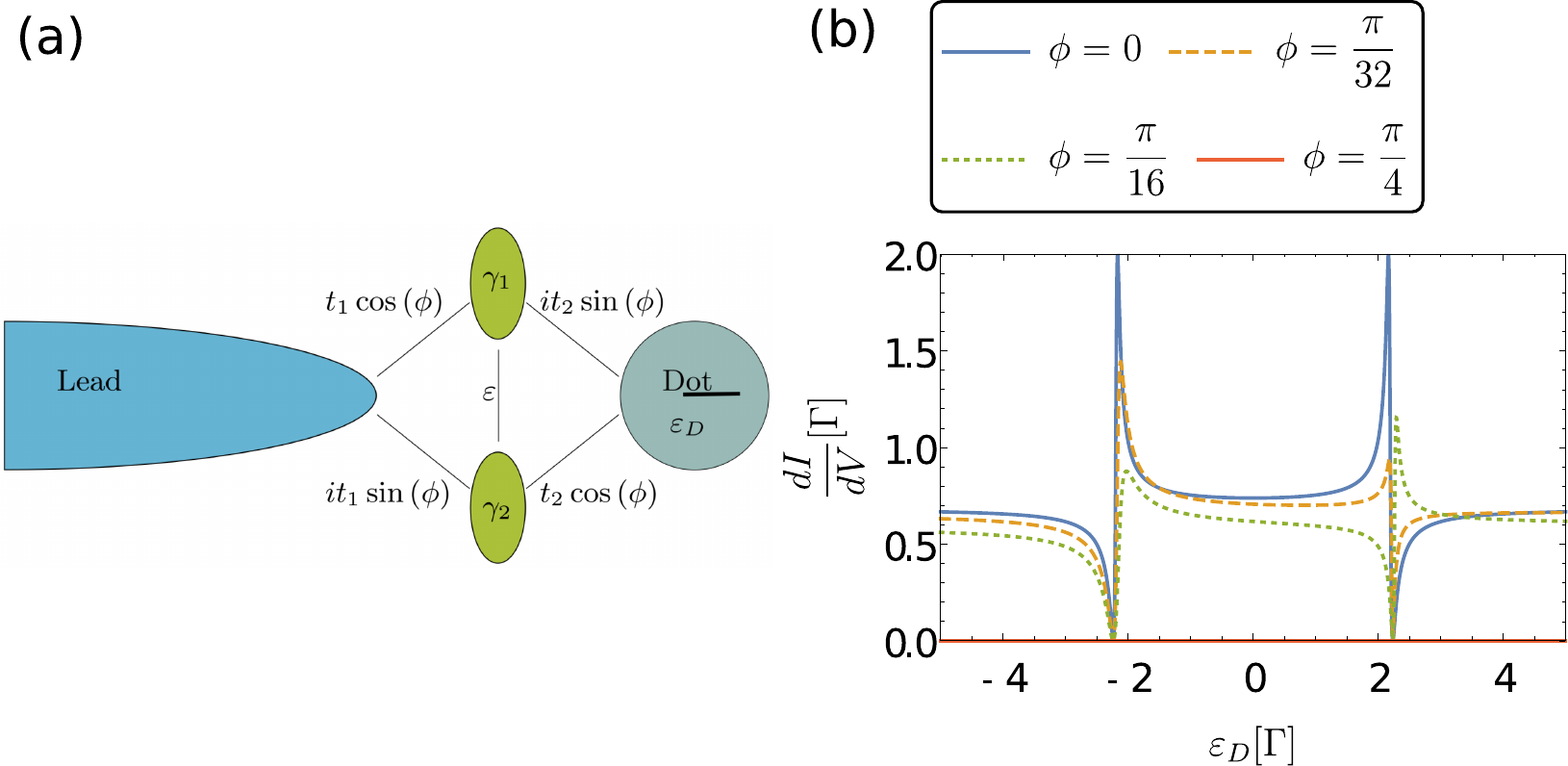}
	\caption{(a) Sketch of the considered setup with tunnel-couplings to both MBSs. The parameter $\phi$ tunes the system between pure Majorana-like ($\phi=0$) and pure Dirac-like ($\phi=\pi/4$) coupling continuously. We choose these values for the couplings because of the spatial symmetry of the considered Kitaev chain. Figure taken from \cite{Schuray2017}. (b) Differential conductance in the setup with nonlocal couplings as a function of dot level energy for various angles of nonlocal couplings. Parameters are $eV=3\Gamma$, $|t_2|=\Gamma$, $\varepsilon=0.4\Gamma$. Figure adapted from \cite{Schuray2017}.}
	\label{fig:nonlocal}
\end{figure}
We consider a finite size Kitaev chain, which is contacted with a normal conducting lead at one side and a quantum dot on the other side. In the topologically non-trivial regime the low energy sector of the Kitaev chain is comprised of two MBSs, described by self-hermitian operators $\gamma_i=\gamma_i^\dagger$, localized mostly at the ends of the chain. Due to the finite size of the chain these two MBSs experience a finite energy splitting $\varepsilon\propto e^{-L/\xi}$, with the length of the chain $L$ and the Majorana decay length $\xi$~\cite{Kitaev2001}. We approximate the lead using a linearized dispersion relation at the Fermi edge. We describe the quantum dot as a single fermionic level with energy $\varepsilon_D$. Because we only consider a spinless system double occupation of the quantum dot is not possible. The MBSs are not perfectly localized at the ends of the TSC, but rather decay exponentially along it. For short chains or large decay lengths (i.e. $L\lesssim\xi$) this implies that the wave function of a MBS reaches the other side of the TSC leading to tunnel couplings to both MBSs as shown in Figure~\ref{fig:nonlocal}~(a). Because of the spatial symmetry of the Kitaev chain we consider the ratios of coupling to the MBSs to be the same for the tunneling to the lead and to the quantum dot. The corresponding Hamiltonian is 
\begin{align}
H=&\ H_L+H_M+H_D+H_T,\\
=&-i\hbar v_F\int dx \psi^\dagger(x)\partial_x\psi(x)+i\varepsilon\gamma_1\gamma_2+\varepsilon_\text{D}d^\dagger d\notag\\
&-i\gamma_1\big[t_1\cos(\phi)\psi+t_1\cos(\phi)\psi^\dagger
+i t_2\sin(\phi)d-it_2\sin(\phi)d^\dagger\big]\notag\\
&-i\gamma_2\big[it_1\sin(\phi)\psi-it_1\sin(\phi)\psi^\dagger
+ t_2\cos(\phi)d+t_2\cos(\phi)d^\dagger\big]\notag,
\end{align}
where $\psi^\dagger(x)$ creates an electron in the lead at position $x$ and $d^\dagger$ creates an electron on the quantum dot. The point-like tunneling to the two MBSs is described with the tunneling amplitudes $t_i$, $v_F$ is the Fermi velocity in the lead and $\phi$ parametrizes the ratio of the tunnel couplings. $\phi=0$ corresponds to the case where lead and quantum dot only couple to one MBS (Majorana-like coupling) while $\phi=\pi/4$ corresponds to a situation where both MBSs couple with the same magnitude to quantum dot and lead, respectively (Dirac-like coupling). The chosen parametrization reflects the spatial symmetry of the Kitaev chain and leads to the same degree of Majorana-like couplings to both lead and dot.\\
Our method of choice for calculating the transport characteristics in this setup is the full counting statistics~\cite{Levitov2004}. Its central element is the cumulant generating function (CGF) which is basically the natural logarithm of the Fourier transform of the probability density of $N$ electrons being transferred through the junction during a large measuring time $\mathcal{T}$. We use the Keldysh technique to calculate the CGF. The details of these calculations can be found in references~\cite{Weithofer2014,Schuray2017}. At zero temperature, the CGF is given by 
\begin{equation}
  \ln\chi(\lambda)=\frac{\mathcal{T}}{2}\int_{-eV}^{eV}\frac{dE}{2\pi\hbar}\ln\big[1+p(E)(e^{-2ei\lambda}-1)\big].
  \label{eq:CGF}
\end{equation}
This CGF describes a binomial distribution. The only process which contributes to transport is Andreev reflection which transfers two electrons through the junction depicted by the two elementary charges $e$ in front of the counting field $\lambda$. These Andreev reflections occur with probability $p(E)$. The differential conductance is simply obtained by taking the derivative of the CGF with respect to the applied bias voltage $V$ and the counting field and then setting the counting field to zero yielding
\begin{equation}
 \frac{dI}{dV}=\frac{2e^2}{h}p(eV).
 \label{didv}
\end{equation}
Also, the current noise and the Fano factor can be readily obtained with this formalism~\cite{Schuray2017}. The differential conductance has local maxima and minima which we call resonances and antiresonances. The resonances in the differential conductance correspond to the spectrum of the system without lead but which are broadened by the tunneling to the lead. In the following, we will discuss the differential conductance in more detail for different degrees of nonlocal couplings.
\subsection{Fano resonances}
\begin{figure}
 \centering
 \includegraphics[width=\textwidth]{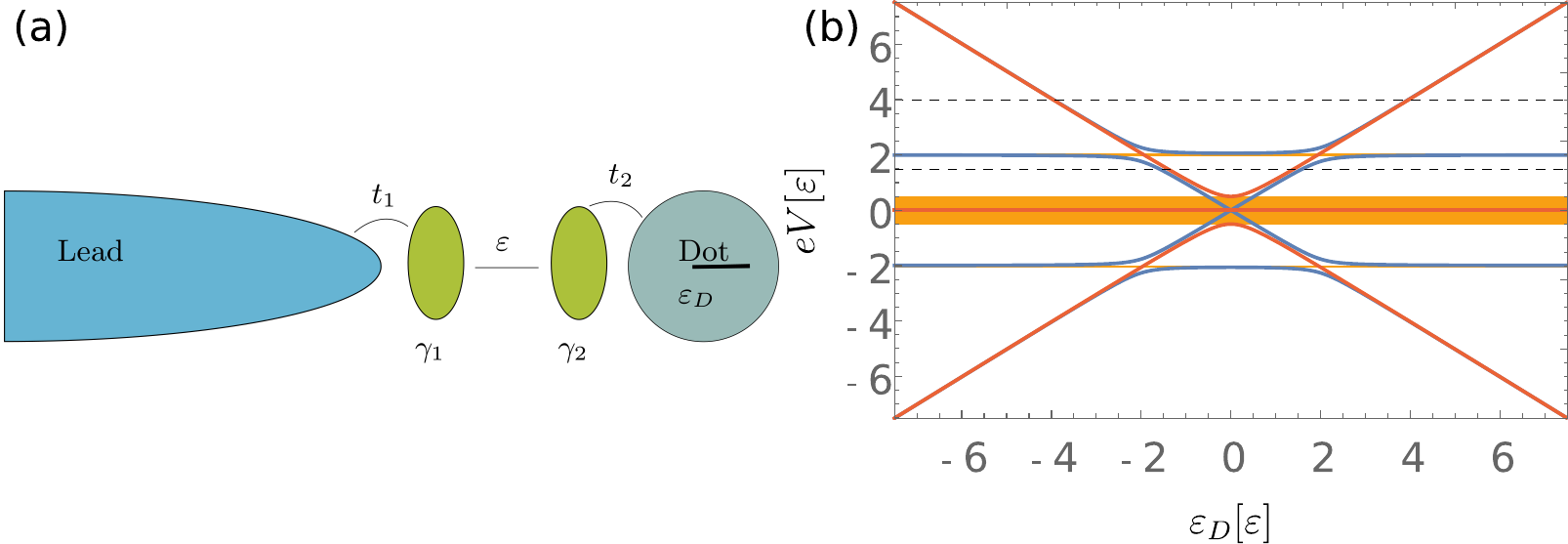}
 \caption{(a) Sketch of the setup under consideration in which a spinless normal conducting lead contacts one of two MBSs $\gamma_1$. The other MBS $\gamma_2$ is contacted by a quantum dot with energy level $\varepsilon_D$. The two MBSs experience a finite energy splitting $\varepsilon$. Figure taken from \cite{Schuray2017}. (b)Resonances ($p=1$, blue) and antiresonances ($p=0$, red) of the differential conductance as a function of applied bias voltage energy $eV$ and quantum dot level energy $\varepsilon_D$. The resonances correspond to the spectrum of the system without the lead. The quantum dot states hybridize with the MBS which leads to an avoided crossing. The areas with an orange background show the parameter space in which no Fano resonances can be found.  Figure adapted from \cite{Schuray2017}.}
 \label{fig:resandanti}
\end{figure}

Now, let us first consider a pure Majorana-like coupling ($\phi=0$). A sketch of this configuration can be seen in Fig.~\ref{fig:resandanti}~(a).
Then the probability for an Andreev reflection $p$ is
\begin{align}
p(eV)=&\ \frac{4\Gamma^2(eV)^2}{\left(\frac{4\varepsilon^2(\varepsilon_{\text{D}}^2-\left(eV\right)^2)}{4|t_2|^2+\varepsilon_{\text{D}}^2-(eV)^2}-(eV)^2\right)^2+4\Gamma^2(eV)^2},
\label{eq:pe}
\end{align}
where $\Gamma=2\pi\rho_0|t_1|^2$. Here, $\Gamma/\hbar$ is the tunneling rate between the lead and $\gamma_1$ with the density of states $\rho_0=(2\pi\hbar v_F)^{-1}$. The differential conductance shows resonances which correspond to perfect Andreev reflection ($p=1$) and antiresonances which correspond to the blocking of transport ($p=0$). As shown in Figure~\ref{fig:resandanti}~(b) the resonances correspond to the spectrum of the system without the lead. The resonances at $\varepsilon_D\approx2\varepsilon$ show an anticrossing which corresponds the hybridization of the quantum dot state with the MBS.\\
It can be shown that the Fano-Beutler formula~\cite{Miroshnichenko2010} can be used to describe the differential conductance as function of quantum dot level energy at fixed bias in certain parameter regimes~\cite{Schuray2017}. These Fano resonances (FR) emerge when a continuous path interferes with a resonant path~\cite{Fano1935}. The first path corresponds to a direct Andreev reflection of an electron or hole which is a continuous path with respect to the quantum dot level energy $\varepsilon_D$. The second path includes an Andreev reflection where the incoming electron or hole first passes through the two MBSs, then virtually occupies the quantum dot before entering the Cooper pair condensate. Here, the second described path is of course resonant with the quantum dot level energy.\\
The differential conductance shows two FRs which are mirrored at $\varepsilon_D=0$ (see Fig.~\ref{fig:nonlocal}~(b), $\phi=0$). The asymmetry parameter of a single FR $q=\pm((eV/2)^2-\varepsilon^2)/(\Gamma(eV/2))$ has a different sign for the resonance at negative $\varepsilon_D$ than for the one at positive $\varepsilon_D$. By tuning the applied voltage bias energy through the Majorana overlap energy the sign of the asymmetry parameter changes for both resonances. The position of the FRs is mostly determined by the applied bias voltage. 
Furthermore, we want to emphasize the relation between the two Fano resonances at fixed bias voltage. They are completely symmetric with respect to each other which we directly attribute to the Majorana nature of the state that the quantum dot is coupled to. Because the MBS is an electron-hole symmetric particle, electron and hole degrees of freedom (positive and negative $\varepsilon_D$) of the quantum dot couple in the same way leading to the above mentioned symmetry. Fano resonances also appear due to interfering transport channels in Cooper pair splitters \cite{Fulop2015,Dominguez2016}. However, in these conventional hybrid systems, the above mentioned symmetry of Fano resonances is absent. \\
Next, we allow for a coupling to both MBSs. In Figure~\ref{fig:nonlocal}~(b), we show the resulting differential conductance as function of $\varepsilon_D$ for various $\phi$. For $\phi=\pi/4$, the coupling to both MBSs is equivalent to coupling to a single Dirac fermion. This leads to a blocking of Andreev reflection as this Dirac fermion has no superconducting properties.\\
In the parameter range $0<\phi<\pi/4$ there is a mixing between Majorana-like and Dirac-like couplings, which destroys the signature of the electron-hole symmetry of the MBS and leads to an asymmetric hybridization of the electron and hole states on the quantum dot with the MBSs. Because of this the two arising FRs also loose their symmetric property and, depending on the system parameters, the asymmetry parameter of the two FRs can even have the same sign which is not the case for $\phi=0$ (c.f. Fig~\ref{fig:nonlocal}, $\phi=\pi/16$). In general, this shows that not only the quantized differential conductance of $2e^2/h$ is a clear indicator of the MBS (it is not quantized for $\phi\neq0$), but also the symmetry $dI/dV(\varepsilon_D)=dI/dV(-\varepsilon_D)$ is a signature which we can directly attribute to the nature of an isolated MBS.\\
So far, to the best of our knowledge, the proposed setup was not yet realized experimentally. However, there have been reports on experiments that focus on the coupling between a quantum dot and a Majorana nanowire~\cite{Deng2016,Deng2018}. In contrast to our suggested setup the electronic lead in the experiments was coupled to the quantum dot and not directly to the Majorana nanowire. The differential conductance measurement in the first experiment~\cite{Deng2016} showed asymmetric hybridizations between the dot states and the low energy in gap states. Several groups pointed out that this can be explained with the coupling to two MBSs~\cite{Schuray2017,Prada2017,Clarke2017}. The second publication~\cite{Deng2018} focused on the analysis of this coupling to a second MBS and these authors measured a quality factor $q=1-\eta^2=1-\tan^2\phi$ which reflects the ``Majorananess'' of the coupling~\cite{Prada2017, Clarke2017, Penaranda2018}. They were able to achieve quality factors of up to $q=0.97$ which corresponds to nearly isolated MBSs. The advantage of our proposed setup over the experimentally realized setups~\cite{Deng2016,Deng2018} is that lead and dot couple to the nanowire on different sides. This allows us to probe non-local properties.

\section{$s$-wave-$p$-wave Josephson junction}
\label{Sec:spincanting}
In the previous section, we have shown that couplings due to the not completely localized wave function of a MBS can be probed with a quantum dot and a normal conducting lead. In the following section, we analyze these couplings by using a superconducting lead and focus on the resulting equilibrium Josephson current. The equilibrium Josephson current is mediated by a phase bias between the two superconductors without any applied voltage bias. We therefore consider a JJ consisting of an $s$-wave superconducting lead and a finite size semiconductor-superconductor hybrid nanowire in the topologically non-trivial phase giving us access to both MBSs via the finite decay length of their wave functions~\cite{Prada2017}. In contrast to a TSC-TSC JJ the energy phase relation shows only a $2\pi$ periodicity.\\
Similiar setups have been already considered by other authors, however, their works either focus on non-equilibrium transport~\cite{Peng2015,Sharma2016,Zazunov2016,Setiawan2017,Setiawan20172} or infinite size TSCs~\cite{Zazunov2012,Ioselevich2016,Zazunov2018}.
It was suggested that for a pure $s$-wave-$p$-wave JJ the equilibrium Josephson current is blocked~\cite{Zazunov2012}. However, for a nanowire system with semi-infinite wires it was already shown that a residual $s$-wave pairing of the non-Majorana states contribute a finite supercurrent~\cite{Ioselevich2016,Zazunov2018}. Moreover, it was shown that a coupling from an $s$-wave lead to two MBSs with non-parallel spins in two different TSCs~\cite{Zazunov2017} or the coupling to Majorana Kramers pairs in time reversal invariant TSCs~\cite{Schrade2018} can result in a finite supercurrent.\\
In contrast, we show that in a \textit{single} finite size Majorana nanowire with broken time reversal symmetry the MBSs can provide a supercurrent if, and only if, both MBSs couple to the superconducting lead. We use a low energy approach and quasi-degenerate perturbation theory to derive an effective Hamiltonian in order to calculate the ground state Josephson current.
We also show that the critical current is governed by the spin canting angle difference of the Majorana wave function at the tunnel-interface. For completeness, we include a numerical treatment of the model and find the high-energy contributions due to the residual $s$-wave pairing as well as contributions attributed to the MBSs and suggest an experimental scheme to separate these two contributions.
\subsection{Model}

\begin{figure}
 \centering
 \includegraphics{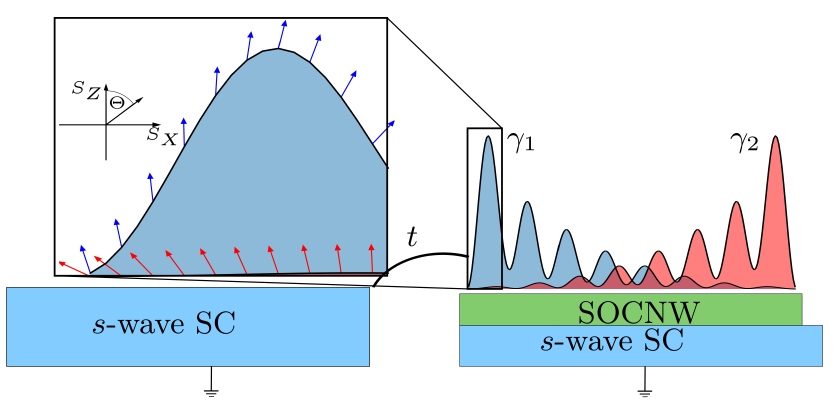}
 \caption{Sketch of the considered $s$-wave-$p$-wave JJ with calculated wave functions of the two MBSs $\gamma_1$ and $\gamma_2$. The arrows in the enlarged window correspond to the spin directions of the corresponding MBS wave functions and rotate with the position within the wire. The electron tunneling takes place between the two superconductors with tunneling amplitude $t$ creating overlaps with both MBSs. Figure taken from \cite{Schuray2018}.}
 \label{fig:setup2.png}
\end{figure}
The Hamiltonian of the system under consideration can be split into three parts
\begin{equation}
 H=H_{BCS}+H_{NW}+H_T,
 \label{eq:spham}
\end{equation}
where $H_{BCS}$ describes the superconducting $s$-wave lead, $H_{NW}$ describes the Majorana nanowire and $H_T$ specifies the electron tunneling between the lead and the TSC. We treat the Majorana nanowire using the Bogoliubov-de Gennes formalism~\cite{Lutchyn2010,Oreg2010}
\begin{equation} 
	H_{NW}=\frac{1}{2}\int_0^L \Psi^\dagger(x) \mathcal{H}_{BdG}^{NW} \Psi(x) dx, 
	\label{NWBdG}
\end{equation}
where $L$ is the finite length of the nanowire and $\Psi^\dagger(x)=\left[\psi_\uparrow^\dagger(x),\psi_\downarrow^\dagger(x),\psi_\downarrow(x),-\psi_\uparrow(x)\right]$ is the Nambu basis creation operator, where $\psi_\sigma^\dagger(x)$ creates an electron with spin $\sigma$ at position $x$ in the nanowire. The Hamiltonian in Nambu basis is then given by
\begin{equation}
 \mathcal{H}_{BdG}^{NW}=\left[\left(-\frac{\hbar^2}{2m^*}\partial_x^2 -\mu\right)-i \alpha \partial_x \sigma_y\right]\tau_z+V_Z\sigma_z+\Delta\tau_x.
	\label{NW}
\end{equation}
This Hamiltonian describes the semiconductor nanowire using an effective mass $m^*$, a chemical potential $\mu$ and the Rashba parameter $\alpha$. Superconductivity is induced by a parent superconductor with $s$-wave pairing in close proximity to the nanowire and the applied Zeeman field $V_Z$ drives the subsystem into a topologically non-trivial phase for $V_Z>\sqrt{\mu^2+\Delta^2}$ in which MBSs emerge at the ends of the wire~\cite{Lutchyn2010,Oreg2010}.\\
The superconducting lead is described using a standard $s$-wave BCS Hamiltonian
\begin{equation}
H_{BCS}=\sum_{\mathbf{k}\sigma}\xi_k c_{\mathbf{k}\sigma}^{\dagger}c_{\mathbf{k}\sigma}+\sum_\mathbf{k} \Delta_{BCS} (c_{\mathbf{k}\uparrow}^\dagger c_{-\mathbf{k}\downarrow}^\dagger+c_{-\mathbf{k}\downarrow}c_{\mathbf{k}\uparrow}),
	\label{HBCS}
\end{equation}
with the superconducting pair amplitude $\Delta_{BCS}$ and the single particle energy of the normal state $\xi_k=\varepsilon_k-\mu$~\cite{Tinkham2004}. Here, the operator $c_{\mathbf{k}\sigma}^\dagger$ creates an electron with momentum $\mathbf{k}$ and spin $\sigma$. For simplicity, we choose a gauge in which the superconducting phase difference between the superconductors $\varphi$ is included in the tunnel Hamiltonian
 \begin{equation} 
	 H_T=\sum_{\mathbf{k}\sigma} te^{i\frac{\varphi}{2}} c_{\mathbf{k}\sigma}^\dagger \psi_{\sigma}(0)+h.c.,
	\label{HT}
\end{equation}
with a momentum and spin independent tunneling amplitude $t$.
\subsection{Low-energy approach}
\begin{figure}
 \centering
 \includegraphics[width=\textwidth]{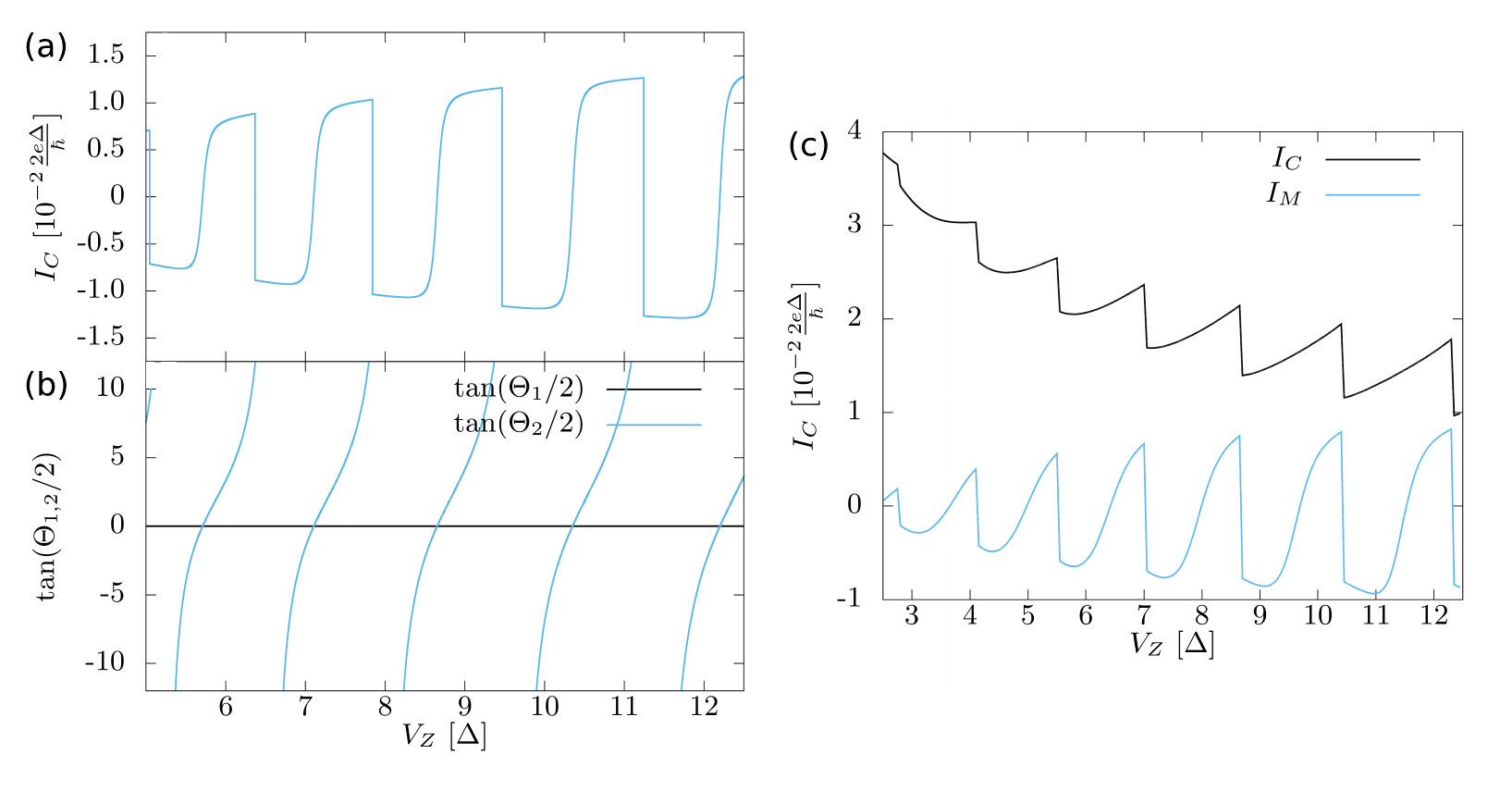}
 \caption{(a) Critical current ($I_C(\varphi=\pi/2)$) in the ground state of the low-energy model and (b) tangent of the spin canting angles at position $x=0$ of the two MBSs (1,2) as a function of Zeeman field $V_Z$. (c) Numerically calculated critical current $I_C$ (black) and Majorana contribution $I_M$ (blue) as a function of applied Zeeman field. The Majorana contribution was extracted following the scheme proposed in the text.The parameters chosen here are $t_S=10\Delta$, $\Delta_{BCS}=\Delta$ and $\tilde t=2.96$ meV. The other microscopic parameters of $(a)$, $(b)$ \& $(c)$ are $m^*=0.015m_e$, $\Delta=0.2$ meV, $\mu=0$, $\alpha=20$ meV nm, $\Gamma=0.004$ meV, $L=1.3$ $\mu$m and $N=M=100$. Figure taken from \cite{Schuray2018}.}
 \label{fig:effectiveI}
\end{figure}
In the topologically non-trivial phase the low energy sector of the nanowire is governed by the non-local fermion comprised of the two MBSs described by the Hermitian operators $\gamma_1$ and $\gamma_2$. This reduces the Hamiltonian in the low energy sector to
\begin{equation} 
	H_{NW}=i\varepsilon\gamma_1\gamma_2,
	\label{effectiveNW}
\end{equation}
with the splitting energy $\varepsilon\approx\hbar^2k_{F,\text{eff}}\left(e^{2L/\xi}m^*\xi\right)^{-1}\cos\left(k_{F,\text{eff}} L\right)$, where the effective Fermi momentum $k_{F,\text{eff}}$ and the Majorana decay length $\xi$ can both be expressed using the microscopic parameters of the model used in equation~(\ref{NW})~\cite{DasSarma2012}. In the low energy description the electron creation operator in the tunneling Hamiltonian equation~(\ref{HT}) can be approximated as
\begin{equation}
 \psi_\sigma(0)\approx\Lambda_{1\sigma}(0)\gamma_1+\Lambda_{2\sigma}(0)\gamma_2,
 \label{eq:create}
\end{equation}
using the electronic components of the Majorana spinor wave functions $\Lambda_{n\sigma}$. We approximately calculate $\Lambda_{n\sigma}$ by considering MBSs which emerge in a semi-infinite wire and then cut the wave functions off at position $L$. The details of this calculation and the expression of $\varepsilon$ in the microscopic parameters can be found in reference~\cite{Schuray2018}.\\
In order to derive a full effective low-energy Hamiltonian we resort to quasi-degenerate perturbation theory with respect to the tunneling Hamiltonian $H_T$~\cite{Schuray2018}. 
To second order in $H_T$ and for $\Delta_{BCS}\gg\varepsilon$ the eigenenergies of the JJ are
\begin{equation}
	\label{eq:Heffeo}
E_{o(e)}^{(2)}=\mp\varepsilon\pm 2\pi \nu(0)\left(it^2\left[\Lambda_{1\uparrow}(0)\Lambda_{2\downarrow}(0)-\Lambda_{1\downarrow}(0)\Lambda_{2\uparrow}(0)\right]e^{i\varphi}+h.c.\right).
\end{equation}
Here, $\nu(0)$ is the density of states in the lead at the Fermi level. Because only Cooper pairs can be transferred across the junction the parity in the nanowire is a conserved quantity and the lower signs of equation~(\ref{eq:Heffeo})  correspond to the odd ($o$) parity while the upper signs correspond to the even ($e$) parity states of the JJ. The MBSs wave functions enter equation ~(\ref{eq:Heffeo}) in spin singlet form, which reflects the fact that the Cooper pairs in the lead are spin singlets.\\
Because the Majorana wave functions have no component in spin $y$-direction~\cite{Sticlet2012} we can parametrize it as
\begin{equation}
 	\begin{pmatrix}
		\Lambda_{n\uparrow}(0)\\
		\Lambda_{n\downarrow}(0)
	\end{pmatrix}=i^{n-1}\kappa_n\left(\begin{matrix}
			\cos(\Theta_n/2)\\
			\sin(\Theta_n/2)
	\end{matrix}\right),
\end{equation}
where $\kappa_n$ is real valued and $\Theta_n$ is the spin canting angle of the $n$-th MBS at the junction. In general, the wave function should vanish at the end of the nanowire (boundary condition), but in order to include the wave function in our point like tunneling model we treat $\kappa_1$ as a free parameter and because of the exponential spatial decay of the MBS we approximate $\kappa_2=\kappa_1\exp\left(-L/\xi\right)$. The spin canting angle of the two MBSs can be different at the interface $x=0$, because the spin of the second MBS rotates as its wave function traverses the nanowire. If we insert this parametrization into equation~(\ref{eq:Heffeo}), we find
\begin{equation}
	E _{e(o)}^{(2)}=\mp\varepsilon\pm\Gamma\cos(\varphi)\sin\left(\frac{\Theta_1-\Theta_2}{2}\right)e^{-L/\xi},
	 \label{HEO}
 \end{equation}
where $\Gamma=4\pi\kappa_1^2t^2\nu(0)$ and we choose $t$ to be real.
Due to the fact that the Cooper pairs in the $s$-wave leads are spin singlets, they can only be transferred across the junction if the spins of the MBSs point in different directions and an optimal Cooper pair transport is achieved for an anti-parallel spin configuration, which is reflected by the sine dependence of the spin canting angle difference in the $\varphi$-dependent part of the eigenenergies.\\
The equilibrium supercurrent is then obtained by taking the derivative of the ground state energy with respect to $\varphi$
\begin{align}
	I(\varphi)=\frac{2e}{\hbar}\partial_\varphi\text{min}(E_{e}^{(2)},E_{o}^{(2)})=I_C(\varphi)\sin(\varphi),
	\label{CPR}
\end{align} 
where the $\varphi$-dependence of the critical current $I_C$ only results in a sign change if the ground state changes between even and odd parity as function of $\varphi$. As seen in Figure~\ref{fig:effectiveI}~(a) the critical current is oscillating as function of applied Zeeman field with jumps at points where the parity of the ground state changes. The oscillation amplitude is growing with increasing Zeeman field because of a larger decay length for higher Zeeman fields. The oscillations can be traced back to the spin canting angle difference. Figure~\ref{fig:effectiveI}~(b) shows that the spin of the first MBS is polarized in the direction of the magnetic field, while the spin of the second MBS rotates at the location of the tunnel junction as the Zeeman field is changed.
\subsection{High energy contributions and tight-binding approach}
In order to underline our analytical findings in the low-energy effective model we consider the equilibrium Josephson effect with the full model described with Eq.~(\ref{NWBdG}). Therefore, we discretize the full Hamiltonian 
\begin{align}
 H=&\sum_{j=1}^N \Psi_j^\dagger \left[(\frac{\hbar^2}{m^*a^2}-\mu)\tau_z+V_Z\sigma_z+\Delta\tau_x\right]\Psi_j+\Bigg\{\Psi_j^\dagger \left[-\frac{\hbar^2}{2m^*a^2}\tau_z+i\frac{\alpha}{a}\tau_z\sigma_y\right]\Psi_{j-1}\notag\\
 &+\sum_{j=1}^M\left\{t_{S} \sum_\sigma c_{j,\sigma}^\dagger c_{j-1,\sigma}+\Delta_{BCS}e^{i\varphi}c_{j,\uparrow}c_{j,\downarrow}\right\}+\tilde{t} c_{M,\sigma}^\dagger\psi_{1,\sigma}+h.c.\Bigg\},
\end{align}
where $\Psi_j^\dagger$ is the creation operator in Nambu basis in the Majorana nanowire, while $c_j^\dagger$ creates an electron in the $s$-wave lead. In order to keep the length of the nanowire fixed we introduce $a=L/N$ with $N$ being the number of lattice sites in the nanowire. The  lead is also discretized using $M$ lattice sites and the hopping inside the lead $t_S$ can be connected to the bandwidth of the superconductor. The tunneling amplitude $\tilde t$ characterizes the tunneling between the last site of the lead and the first site of the nanowire. We diagonalize the Hamiltonian and calculate the critical current
\begin{equation}
 I_C=\max_\varphi I(\varphi),
\end{equation}
where the suppercurrent $I(\varphi)$ follows from $I(\varphi)=\frac{2e}{\hbar}\partial_\varphi\sum_{E_i<0}E_i(\varphi)$. In Figure~\ref{fig:effectiveI}, we show the critical current and the Majorana contribution to it.
While the Majorana contribution qualitatively confirms the analytical findings, higher energy contributions conceal most of the Majorana features. The higher energy contributions come from a residual $s$-wave pairing in the TSC~\cite{Zazunov2018,Cayao2018}. The critical current still shows jumps at parity crossings, however, in contrast to the Majorana contributions the magnitude of the critical current is reduced with increased Zeeman field. Also, the critical current is never completely blocked regardless of the spin canting of the MBSs.

To experimentally unveil the Majorana contribution from the critical current we suggest an experimental scheme based on quasiparticle poisoning. A quasiparticle poisoning event changes the parity of the low energy sector. Thus, these events should be avoided in topological quantum computing schemes. A quasiparticle poisoning time, the time scale on which these events occur, of $T_P\approx100\mu$s is experimentally feasible~\cite{OFarrell2018}. The scheme is based on  measuring the supercurrent for both parities of the low energy sector spanned by the two MBSs with a fixed set of parameters $\Delta_Z$, $I_C$. Because the background of the high energy contribution is the same for both parities the difference of the two critical currents removes the background and owing to the particle-hole symmetry of the superconductor system all that is left is twice the Majorana contribution $I_M$. In order to realize this scheme the current needs to be measured within $T_P$ with a sensitivity of less than $10^{-2}\frac{2e\Delta}{\hbar}$ in order to resolve the Majorana contributions. This high sensitivity within the short quasiparticle poisoning time is experimentally challenging but feasible.

\section{TSC-TSC junction in silicene}
\label{Sec:silicene}
In the model discussed in the previous section the focus has been placed on a superconducting JJ between a topological superconductor and a non topological superconductor~\cite{Zazunov2018,Zazunov2017,Cayao2018}.
In such a JJ, the current phase relation features a $ 2\pi $ periodicity like the traditional Josephson effect.
However, JJs between two topological superconductors are predicted to feature promising avenues to probe localized Majorana zero modes~\cite{Kitaev2001,Alicea2011,Peng2016,Klees2017,Sticlet2018,Fu2009b,Kane2015, Crepin2014,Pikulin2012,Rainis2012,San-Jose2012,Houzet2013,Virtanen2013,Cayao2017,Cayao2018b,Bocquillon2016,Wiedenmann2016,Bocquillon2018,Dominguez2012,Dominguez2017,Pico-Cortes2017,Deacon2017,Beenakker2013b,Frombach2018}.
In their paper \cite{Fu2009} Fu and Kane predict the existence of MBSs localized inside such a JJ as well as a $ 4\pi $ periodicity of the current phase relation, an effect closely linked to the existence of MBSs which is known as the fractional Josephson effect.
Additionally, it is predicted that including interactions in such JJs can result in the existence of excitations similar to $Z_4$ parafermions linked to an $ 8\pi $ periodicity~\cite{Zhang2014}.
However, these fractional effects are only preserved, as long as the fermion parity, i.e. the number of fermionic particles in the JJ modulo 2, is conserved inside of the junction.
In general, effects breaking this fermion parity conservation are present in real JJs and therefore possibly spoil this experimental signature.
Since these predictions a lot of work has been done to overcome this problem of quasiparticle poisoning and first experimental signatures of the fractional Josephson effect could be observed both in junctions based on nanowires \cite{Rokhinson2012,Laroche2019} as well as junctions based on HgTe \cite{Wiedenmann2016,Deacon2017}.
These experiments focus on dynamical effects of JJs like Shapiro steps and Josephson radiation and are performed out of equilibrium.
However, a scheme to identify a 4$\pi$ Josephson effect based on dc properties of JJs was proposed by Beenakker \textit{et al.}~\cite{Beenakker2013}, even in the presence of slow quasiparticle poisoning.
By looking at long junctions, i.e. JJs where the distance between the two superconductors is large compared to the superconducting coherence length $ \xi_0 $, they found that the critical current of topologically non-trivial JJs is twice as large as the critical current of topologically trivial JJs.
This now calls for a tunable JJ in which one can tune between a topologically trivial and topologically non-trivial JJ to extract the proposed signature.
Such junctions can be provided by making use of the buckled structure of silicene or similar compounds \cite{Frombach2018}, which we will review in the following.

\subsection{Model}

\begin{figure}
\centering
	\includegraphics[width = 0.65\columnwidth]{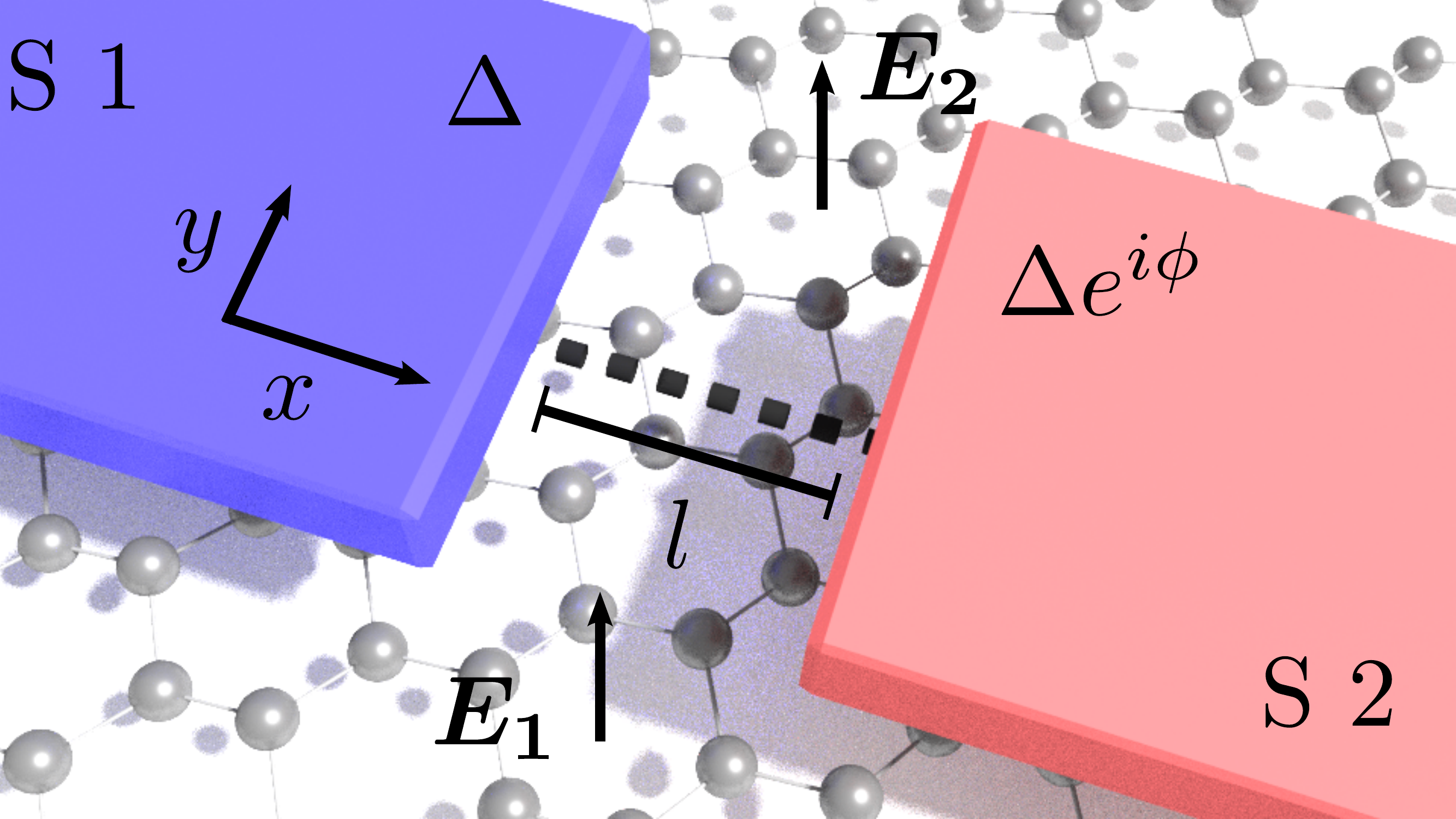}
	\caption{Two superconductors (S~1 and S~2) are placed on top of a sheet of silicene with a distance $ l $ and a phase difference $ \phi $ between the two.
	The two electric fields $ \bm{E}_1 $ and $ \bm{E}_2 $ perpendicular to the silicene sheet define the helical edge states (dashed black line) at the border between the two regions. Figure taken from~\cite{Frombach2018}.}
	\label{fig:silicene:setup}
\end{figure}

To build a JJ that is tunable between a topologically trivial and a topologically non-trivial state we consider a sheet of silicene with proximity induced superconductivity via two superconducting leads (Fig.~\ref{fig:silicene:setup}).
In addition, two electric fields are applied perpendicular to the sheet.
As we will see below, the border between the two regions of different electric fields will define helical edge states mediating the Josephson effect.
Such a system for low energies can be described by the following Hamiltonian
\begin{equation}
	H = \frac{1}{2} \int d^2 x\, \Psi^\dagger(\bm{x}) \mathcal{H}(\bm{x}) \Psi(\bm{x}),
	\qquad
	\mathcal{H}(\bm{x}) = \mathcal{H}_0 + \mathcal{H}_{S} + \mathcal{H}_{I},
\end{equation}
where we use the standard Nambu basis
\begin{equation}
	\begin{aligned}
		\Psi(\bm{x}) &= 
		(\psi_{\uparrow}(\bm{x}), 
		\psi_{\downarrow}(\bm{x}),
		{\bar \psi_{\downarrow}}^\dagger(\bm{x}),
		-{\bar \psi_{\uparrow}}^\dagger(\bm{x}))^T, \\
		\psi_{s}(\bm{x}) &= 
		(c_{A K s}(\bm{x}),
		c_{B K s}(\bm{x}),
		c_{A K' s}(\bm{x}),
		c_{B K' s}(\bm{x}))^T,
	\end{aligned}
\end{equation}
and $ {\bar \psi}_{s}(\bm{x})$ is obtained from $\psi_{s}(\bm{x})$ by the substitution $K\leftrightarrow K'$.
The Hamiltonian can be separated into three parts.
The first part \cite{Kane2005,Ezawa2015,Drummond2012}
\begin{equation}
	\mathcal{H}_0 = -i\hbar v_F (\partial_x \rho_z \tau_z \sigma_x + \partial_y \rho_z \sigma_y) 
	- \Delta_{\textrm{SO}} \rho_z s_z \tau_z \sigma_z
	+ m \rho_z \sigma_z
\end{equation}
describes the electronic properties of a single sheet of silicene.\footnote{
This notation differs from the one in reference~\cite{Frombach2018} by an additional sign in the changed spin basis.
}
The first term containing the Fermi velocity $ v_F $ describes the kinetic energy.
The second term describes the intrinsic spin-orbit interaction $ \Delta_{SO} $ while the last term describes the staggered potential $ m \propto E_z $ resulting from the perpendicular electric field.
The Pauli matrices $ \sigma $~/~$ \tau $~/~$ s $~/~$ \rho $ correspond to the sublattice $ (A, B) $ / valley $ (K, K') $ / spin $ (\uparrow, \downarrow) $ / particle-hole degree of freedom.
The $ s- $wave superconducting proximity effect is described via
\begin{equation}
	\mathcal{H}_{S} = \Delta (\cos(\phi) \rho_x + \sin(\phi) \rho_y),
\end{equation}
where $ \Delta $ is the superconducting pairing gap and $ \phi $ the phase difference between the two superconductors.
Finally, scattering events between the two valleys of silicene are described by
\begin{equation}
	\mathcal{H}_I = \delta \rho_z \tau_x,
\end{equation}
where $ \delta $ is the strength of the intervalley scattering, which we assume is generally present.

\subsection{Chern Number}
\label{subsec:chernNumber}

In order to understand why this system features the described tunability we can first look at a simplified system neglecting superconductivity ($ \mathcal{H}_S $) and intervalley scattering ($ \mathcal{H}_I $) and consider only the silicene sheet in the presence of two electric fields ($ \mathcal{H}_0 $).
This system preserves both the spin as well as the valley subspace so that they can be expressed via the eigenvalues $ \xi, \eta = \pm 1 $ of their respective Pauli z-matrices $ s_z $ and $ \tau_z $.
The resulting energy spectrum
\begin{equation}
	\varepsilon_{\bm{k}} = \pm\sqrt{(\hbar v_F \bm{k})^2 + (m - \eta \xi \Delta_{\textrm{SO}})^2},
\end{equation}
is hyperbolic with an energy gap of $ 2 \times (m - \eta \xi \Delta_{\textrm{SO}}) $.
Because the mass parameter $ m $ is proportional to the $ z $-component of the applied electric field this energy gap closes at a critical electric field $ E = \pm E_c $ with $ E_c = \Delta_{SO} / d \approx 17 $meV/\AA \,\cite{Ezawa2012,Frombach2018} where $ 2d $ is the separation along the $ z $-direction of the two sublattices of silicene.
At which of the two signs of the electric field the gap closes depends on the spin and valley configuration.
By calculating the spin and valley dependent Chern numbers
\begin{equation}
	\label{eq:silicene:chernNumber}
	n_{\eta \xi} = \frac{\eta}{2} \sign(m - \xi \eta \Delta_{\textrm{SO}}),
\end{equation}
we can see that this gap closing is accompanied by a topological phase transition.
While the overall Chern number
\begin{equation}
	n = \sum_{\eta\xi} n_{\eta\xi} = 0
\end{equation}
vanishes due to the entire system being time reversal invariant a topological $ \mathbb{Z}_2 $ invariant
\begin{equation}
	\nu = \sum_{\eta} \frac{n_{\eta \uparrow} - n_{\eta \downarrow}}{2} \mod 2
	= \left\{ \begin{aligned}
	& 0 & \abs{E_z} > E_c \\
	& 1 & \abs{E_z} < E_c
	\end{aligned} \right.
\end{equation}
distinguishes between the topologically trivial ($ \nu = 0 $) and non trivial ($ \nu = 1 $) regime.
In the topologically non-trivial regime the system is a quantum spin Hall insulator featuring a single set of spin-helical edge states at the sample edge.
By applying an electric field stronger than the critical field $ E_c $ perpendicular to the silicene sheet the system can be tuned into the topologically trivial regime which in turn also destroys the topological edge states.

\begin{figure}
\centering
	\includegraphics[width = 0.5\columnwidth]{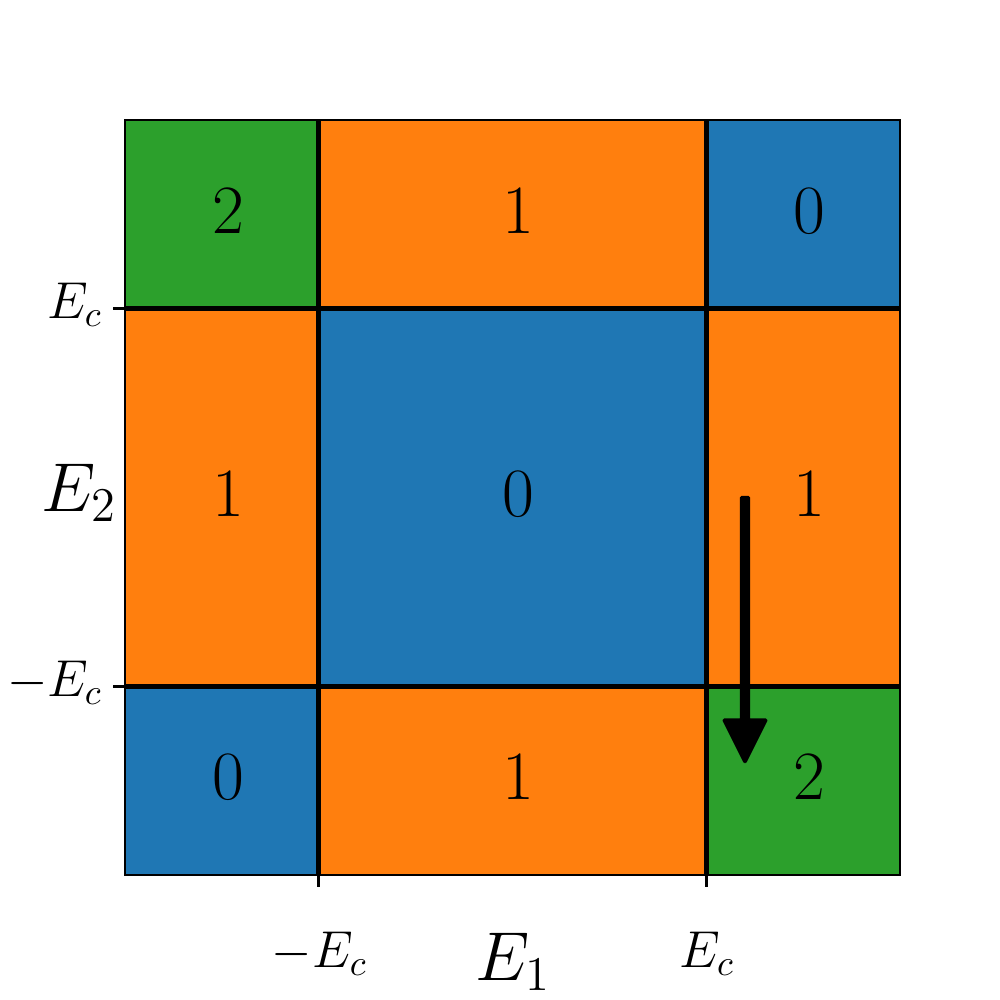}
	\caption{
		Phase diagram of the three different scenarios of edge states forming at the boundary between two regions of a silicene sheet featuring two different electric fields $ E_1 $ and $ E_2 $.
		The numbers indicate the number of spin-helical edge state pairs at the boundary.
		In the green region the two sets of spin-helical edge states form a single set of spin-degenerate edge states.
		The black arrow indicates an example path to tune from spin-helical to spin-degenerate edge states at the border.
	}
	\label{fig:silicene:phaseDiagram}
\end{figure}

With a single applied electric field the system can only either have a single set of spin-helical edge states or no edge states at its sample edge.
A richer situation arises when we look at the boundary between regions with two different electric fields $ E_1 $ and $ E_2 $.
If the spin and valley dependent Chern number~\eqref{eq:silicene:chernNumber} changes across the boundary for a specific spin and valley polarization a topological edge state of corresponding polarization exists running along the boundary between the two regions.
Because the system is time reversal invariant, these edge states always exist in time reversal invariant spin-helical form.
Therefore, the phase diagram in Figure~\ref{fig:silicene:phaseDiagram} lists the number of spin-helical edge state pairs along the boundary while the total number of edge states will always be twice that amount.
By changing the two electric fields three scenarios can be distinguished.

First, if both electric fields are small ($ \abs{E_1}, \abs{E_2} < E_c $) both regions of the silicene sheet are topologically non-trivial ($ \nu = 1 $) and there exist no topological edge states at the boundary (Fig.~\ref{fig:silicene:phaseDiagram}, blue center region).
Also no edge states exist if both electric fields are large and point in the same direction ($ E_1, E_2 > E_c $ or $ E_1, E_2 < -E_c $), however both regions are now topologically trivial ($ \nu = 0 $)(Fig.~\ref{fig:silicene:phaseDiagram}, blue outer corners).

The second scenario features one set of spin-helical topological edge states at the boundary between the two regions (Fig.~\ref{fig:silicene:phaseDiagram}, orange).
This is the case when the electric fields fulfill the condition $ \abs{E_i} < E_c < \abs{E_j} $.
Here one of the two regions is topologically non-trivial ($ \nu = 1 $) while the other is topologically trivial ($ \nu = 0 $).
These edge states are protected from backscattering as they are Kramers partners.

Finally the third scenario features spin-degenerate edge states at the boundary between the two regions which can be separated into two sets of spin-helical edge states (Fig.~\ref{fig:silicene:phaseDiagram}, green).
This scenario occurs when both electric fields are large but point in opposite directions ($ E_i < -E_c $ and $ E_j > E_c $).
In this case both regions are topologically trivial ($ \nu = 0 $).

The existence of edge states in the last scenario does not contradict the fact that the topology does not change across the border,\footnote{
While the $ \mathbb{Z}_2 $ number $ \nu $ does not change across the boundary, individual spin and valley dependent Chern numbers $ n_{\eta\xi} $ do change resulting in the existence of these edge states.
} as these edge states are not topologically protected.
In fact, the existence of backscattering in these spin-degenerate edge states is critical in order to tune a topologically non-trivial JJ to a trivial one.
To achieve the desired tunability to be described in the following section, we will have to tune from a regime with spin-helical (Fig.~\ref{fig:silicene:phaseDiagram}, orange) to a regime with spin-degenerate (Fig.~\ref{fig:silicene:phaseDiagram}, green) edge states.
This can be achieved for instance by first setting $ E_1 > E_c $ while $ E_2 = 0 $ so that spin-helical edge states exist and then tuning $ E_2 < -E_c $ while leaving $ E_1 $ unchanged as indicated by the black arrow in Figure~\ref{fig:silicene:phaseDiagram}.

\subsection{Josephson junction}

With the ability to tune edge states from spin-helical to spin-degenerate we can now turn to JJs mediated by such edge states by including the superconducting term $ \mathcal{H}_S $.
First looking at the case of spin-helical edge states we can project the Hamiltonian $ \mathcal{H} $ onto the subspace spanned by the two edge states and find two Andreev bound states (ABSs)
\begin{equation}
	\Gamma_1 = \int dx \, \varphi(x) \frac{1}{\sqrt{2}} 
	\left[ e^{i\theta} c_{\uparrow}(x) + c_{\downarrow}^\dagger(x) \right],
	\quad
	\Gamma_2 = \int dx \, \varphi(x) \frac{i}{\sqrt{2}} 
	\left[ e^{-i\theta} c_{\downarrow}(x) - c_{\uparrow}^\dagger(x) \right],
\end{equation}
forming inside of the junction by standard wave matching procedures where
\begin{equation}
	\varphi(x) = \sqrt{\frac{\sqrt{\Delta^2 - \varepsilon^2}}{\hbar v_F}}
	e^{- \frac{\sqrt{\Delta^2 - \varepsilon^2}}{\hbar v_F} \abs{x}},
	\qquad
	\theta = \arg \left( \varepsilon + i \sqrt{\Delta^2 - \varepsilon^2} \right)
\end{equation}
and $ c_{s}^{\dagger}(x) $ create an electron in the helical edge state with spin polarization $ s $ at position $ x $ along the boundary.
They feature the $ 4\pi $ periodic energy phase relation
\begin{equation}
	\label{eq:silicene:EPR}
	\varepsilon(\phi) = \pm \Delta \cos \left( \frac{\phi}{2} \right)
\end{equation}
and lie at zero energy if the phase difference across the JJ is equal to $ \pi $.
Linear combinations of these ABSs at zero energy in turn form MBSs
\begin{equation}
	\gamma_1 = \left. \frac{(\Gamma_1 + \Gamma_2)}{\sqrt{2}} \right|_{\varepsilon = 0},
	\qquad
	\gamma_2 = \left. \frac{i(\Gamma_1 - \Gamma_2)}{\sqrt{2}} \right|_{\varepsilon = 0}
\end{equation}      
due to the relation $ \Gamma_1^\dagger|_{\varepsilon = 0} = \Gamma_2|_{\varepsilon = 0} $.

This JJ is topologically non-trivial as it is being mediated by spin-helical topological edge states and features two MBSs and a $ 4\pi $ periodic energy phase relation.
As we tune the electric fields to the regime featuring spin-degenerate edge states the system effectively gets doubled.
Repeating the steps described above reveals that in this situation there indeed exist 4 ABSs with the same energy phase relation~\eqref{eq:silicene:EPR}, where each branch is now two fold degenerate.
However, the crossing at zero energy is no longer protected, as backscattering in the spin-degenerate edge states is no longer prohibited by time reversal symmetry.
This means, that including the time reversal invariant intervalley scattering $ \mathcal{H}_I $ the $ 4 $ ABS can now couple, open a gap in the spectrum at zero energy and turn the energy phase relation 
\begin{equation}
	\varepsilon(\phi) = \pm \sqrt{ \Delta^2 \cos^2 \left( \frac{\phi}{2} \right) + \delta^2 }
\end{equation}
$ 2\pi $ periodic.
As no zero energy states exist in this case there are no longer any MBSs in this system and the JJ is topologically trivial.

As intervalley scattering $ \mathcal{H}_I $ is allowed by the symmetry constraints of the system it will generically be present and can arise for instance through disorder \cite{Wang2014}. A way to tune $\delta$ in a single setup is proposed in \cite{Frombach2018}. By tuning the electric fields along the path depicted in Figure~\ref{fig:silicene:phaseDiagram} we can therefore electrically tune between a fractional JJ and a nonfractional JJ which can potentially be used to demonstrate the topological nature through a dc measurement of the critical current described in \cite{Beenakker2013}.
The topological Josephson effect studied here differs from previous studies \cite{Kwon2004,Tanaka1997,Kashiwaya2000}
by the fact, that spin-degenerate topological edge states, albeit not being protected, exist connecting the superconductors resulting in a conventional Josephson effect.
While the analytical model is based on the short junction limit $ l \ll \xi_0 $ numerical tight binding simulations have shown, that this effect persists into the long junction regime \cite{Frombach2018}.

\section{Corbino geometry topological Josephson junction}
\label{sec4:Corbino}

Several theoretical extensions have been devoted to the search for MBSs in a circular geometry due to the geometrical versatility for the manipulation and braiding of MBSs~\cite{Fu2008,Grosfeld2011,Lindner2012,Park2015,Li2016}. In this section, we consider a Corbino geometry JJ deposited on the surface of a three-dimensional TI \cite{Park2015}, and discuss how to create and braid MBSs, and how to detect the Majorana exchange statistics in an electrical transport experiment.

\subsection{Vortex-bound Majorana fermions}
\label{sec4:s-ti}

A three-dimensional TI is a material characterized by an insulating bulk in coexistence with metallic surface states as a consequence of strong spin-orbit coupling~\cite{Fu2007,Moore2007,Roy2009}. The surface of the TI is described by a Dirac Hamiltonian $H_{\textrm{\tiny TI}} = v_{F} (p_x \sigma_x + p_y \sigma_y)$ in the $x$-$y$ plane~\cite{Hasan2010} where $v_{F}$ is the Fermi velocity and the $\sigma$ Pauli matrices denote spin. Similar to any other metal, the surface becomes superconducting when it is proximity coupled to an $s$-wave superconductor, which can be modeled by the BdG equation
\begin{eqnarray}
& H_{\textrm{\tiny SC/TI}} \, \Psi (\mathbf{r}) = E \, \Psi (\mathbf{r}), \label{sec4:BdGEq} \\
& H_{\textrm{\tiny SC/TI}} = 
\left(
\begin{array}{cc}
H_{\textrm{\tiny TI}} - \mu & \Delta \\
\Delta^{*} & -H_{\textrm{\tiny TI}} + \mu
\end{array} \label{sec4:BdGH}
\right),
\end{eqnarray}
where $\Psi (\mathbf{r}) = (u_{\uparrow},u_{\downarrow},v_{\downarrow},-v_{\uparrow})^{T}$ is the Nambu spinor. This proximity effect is a necessary but not sufficient condition for creating MBSs because the surface remains gapped if there is no vortex in the superconductor. We discuss below how a vortex modifies the order parameter $\Delta$ and binds a MBS. 

\begin{figure}
	\centering
	\resizebox{0.9\columnwidth}{!}{\includegraphics{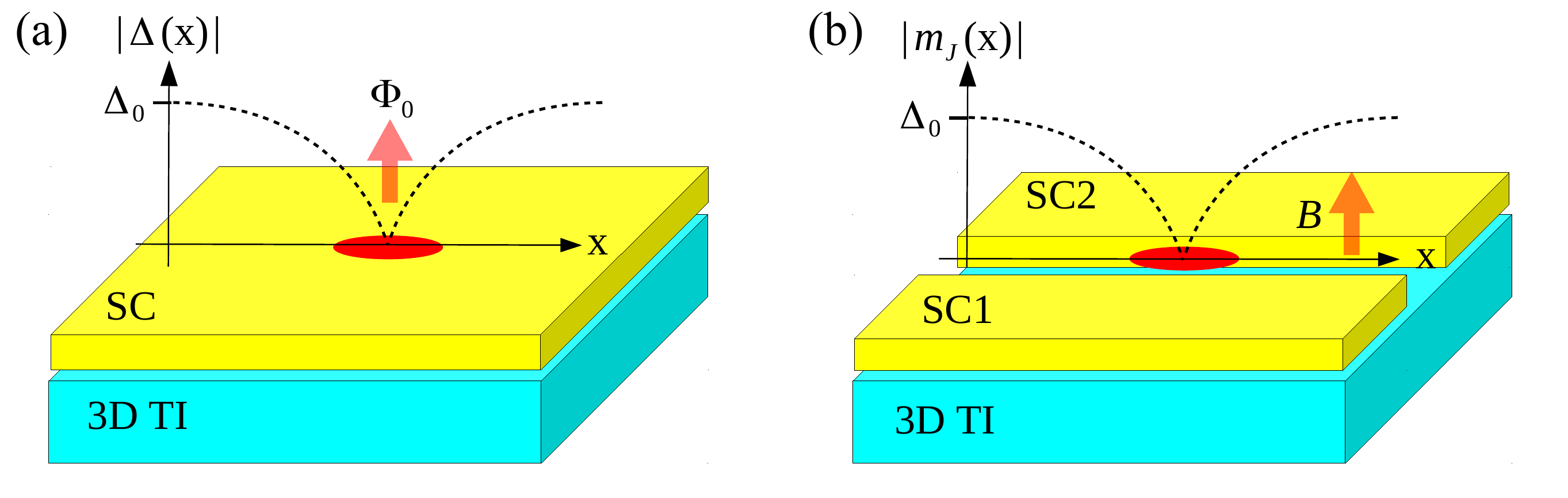}}
	\caption{A Majorana fermion bound to (a) an Abrikosov vortex or (b) a Josephson vortex in a superconductor-topological insulator heterostructure forming a JJ.}
	\label{sec4:fig1}       
\end{figure}

As a consequence of the physical requirement that the superconducting order parameter is a single-valued function, an Abrikosov vortex carries a flux quantum $\Phi_0 = h/(2e)$ and has an order parameter of the form
\begin{equation}
\Delta(r,\theta) = \Delta(r) e^{i\phi-i n_v \theta},\label{sec4:vortex}
\end{equation}
where polar coordinates $(r,\theta)$ are defined with respect to the center of the vortex, $\phi$ is the superconducting phase which is uniform in space. $\Delta(r)$ vanishes at $r=0$ and converges to a constant $\Delta_0$ at large $r$. $n_v=1$ is the winding number of the vortex, which is essential ingredient for the creation of MBSs. It is known that the proximity effect of this order parameter on the TI surface supports a single MBS at zero energy \cite{Fu2008}, as illustrated in Figure~\ref{sec4:fig1}(a). However, it is not easy to find an analytical solution of the Majorana wave function for an arbitrary value of the chemical potential $\mu$. One can prove analytically the existence of a MBS in two limiting cases of $\mu$: $\mu\gg \Delta_0$ and $\mu=0$. For the former case, it is possible to perform an unitary transformation that maps the topological insulator-superconductor heterostructure into a spinless $p_x+i p_y$ superconductor~\cite{Fu2008, Alicea2012}, and for the latter, solving explicitly the BdG equation at $E=0$, the Bogoliubov quasiparticle operator for the zero-energy solution of the vortex is given by  
\begin{equation}
\gamma = \int d^2 r \, e^{-\frac{1}{\hbar v_F}\int^{r}_0 d s \Delta(s)} 
\left[ e^{i \phi/2 + i \pi/4} \psi_{\downarrow}(\mathbf{r}) +
e^{-i \phi/2 - i \pi/4} \psi^{\dagger}_{\downarrow}(\mathbf{r}) \right],
\end{equation}
satisfying $\gamma = \gamma^{\dagger}$, where $\psi_{\sigma}(\mathbf{r})$ is the electron field operator with spin $\sigma$. Note that, in this case, there is no zero-energy solution for the spin-up sector. Recent experiments have reported the experimental signatures of the vortex-bound Majorana state using scanning tunneling spectroscopy~\cite{Nadj-Perge2014,Xu2015,Sun2016,Pawlak2016,Chevallier2016,Wang2018,Liu2018,Menard2019}. 

A Josephson vortex formed in a SC-TI-SC junction with an external magnetic field $B$ applied perpendicular to the TI surface, illustrated in Fig.~\ref{sec4:fig1}(b), can also be used to create MBSs, due to its topological equivalence with the Abrikosov vortex discussed above. If the distance $d$ between the superconductors is small compared to the coherence length and the electromagnetic field produced by the Josephson current (screening effect) is negligible~\cite{Tinkham2004}, subgap states of the JJ can be described by counter-propagating one-dimensional channels along the junction which are coupled by an effective mass $m_J(x) = \Delta_0\, \textrm{cos} [\delta(x)/2]$, where $\delta(x)= 2 \pi B d x/\Phi_0 + \delta_0$ is the local superconducting phase difference along the axis $x$ parallel to the junction~\cite{Fu2008,Potter2013}. The constant parameter $\delta_0$ is determined by the phase difference of the superconductors in the absence of the magnetic field. Here, one can define the position $x=x_v$ at which $\delta(x_v) = \pi$ is satisfied, as the center of the Josephson vortex. Like the center of the Abrikosov vortex, the mass gap vanishes at the center, $m_J(x_v)=0$, and the superconducting order parameter winds around a closed loop encircling the center of the Josephson vortex. The usefulness of the Josephson vortex is that the position $x=x_v$ where a MBS is localized can be controlled by changing the superconducting phase difference $\delta_0$.  

To proceed, we give a brief introduction to braiding of several MBSs. A well separated pair of MBSs, $\gamma_1$ and $\gamma_2$, can be described by non-local fermionic operators, $c = (\gamma_1+i \gamma_2)/2$ and $c^{\dagger}$, whose number operator is $\hat{n} = c^{\dagger} c$ defining a degenerate two states, an even-parity state $|0\rangle$ with $n=0$ and an odd-parity state $|1\rangle$ with $n=1$. An exchange process of $\gamma_1$ and $\gamma_2$ leads to the development $\gamma_1 \rightarrow -\gamma_2$ and $\gamma_2 \rightarrow \gamma_1$, which corresponds to a unitary operation $\gamma_i \rightarrow B_{12}\gamma_i B_{12}^{\dagger}$ with $B_{12}=\exp[(\pi/4)\gamma_1\gamma_2]=[1+\gamma_1\gamma_2]/\sqrt{2}$, where the convention is $\{\gamma_i,\gamma_j\}=2\delta_{ij}$~\cite{Ivanov2001}. In the occupation number basis $\{|0\rangle,|1\rangle\}$, $B_{12}=\exp[i(\pi/4)\sigma_z]$, which produces just a phase factor in the fermion basis $|0\rangle \rightarrow e^{i\frac{\pi}{4}}|0\rangle$, $|1\rangle \rightarrow e^{-i\frac{\pi}{4}}|1\rangle$. To use the full power of non-abelian braiding (i.e. a state change in the manifold of degenerate ground states) one needs at least 4 MBSs, i.e. 2 fermions with basis $\{|ij\rangle\}$, $i,j=0,1$~\cite{Sarma2015}. In the subspace of a given total fermion parity (e.g. even parity with basis states $\{|00\rangle, |11\rangle\}$) braiding MBS 2 and 3 would correspond to a rotation around the $x$-axis
that turns the basis states into a superposition state (e.g. $|00\rangle \rightarrow (|00\rangle+|11\rangle)/\sqrt{2}$~\cite{Ivanov2001,Leijnse2012,Aguado2017}). In the following subsections, we discuss how to realize and detect the exchange of 2 MBSs in a Corbino geometry JJ. For the extension of this proposal to 4 MBSs, we refer readers to reference~\cite{Park2018}. 

\subsection{Creation and braiding of MBSs}
\label{sec4:braiding}

\begin{figure}
	\centering
	\resizebox{0.9\columnwidth}{!}{\includegraphics{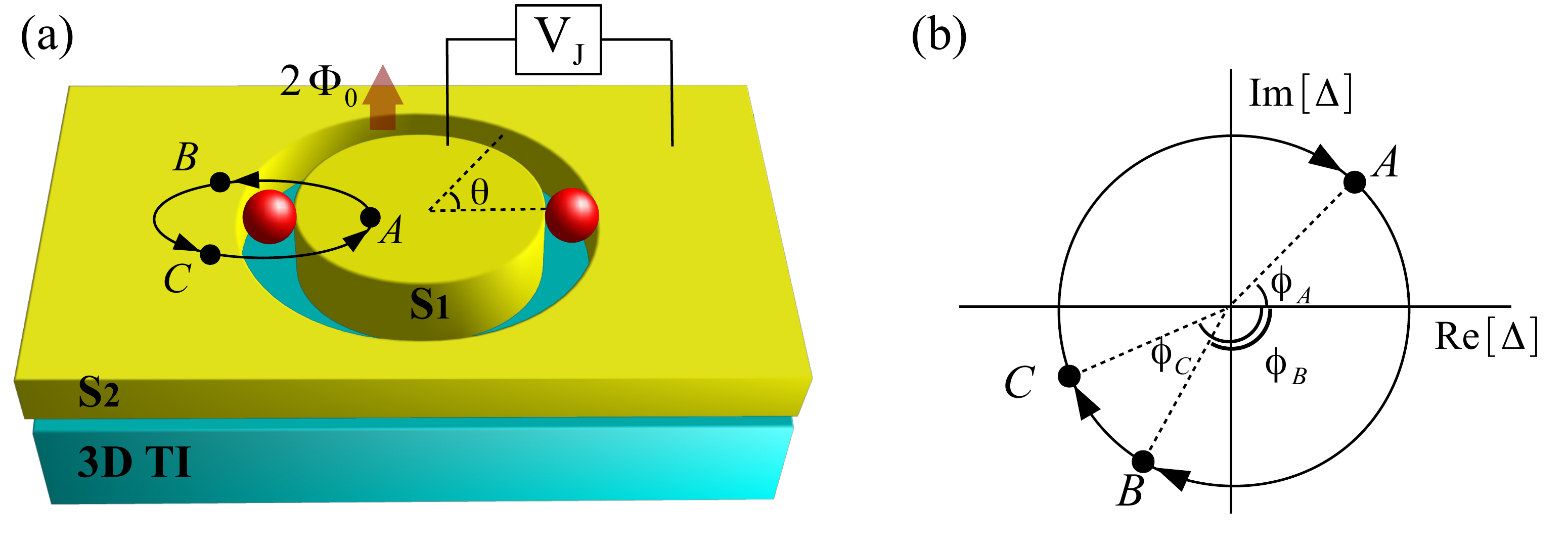}}
	\caption{(a) Schematic of a Corbino geometry topological JJ formed by thin-film superconductors denoted as $S_1$ and $S_2$ which are deposited on the surface of a topological insulator (3D TI). In the presence of two flux quanta, $2 \Phi_0$, two zero-energy MBSs (red balls) form at opposite sides of the junction. Their positions can be moved by applying a small voltage $V_J$ across the junction, enabling us to perform an adiabatic exchange. Figure adapted from~\cite{Park2015}. (b) A winding of superconducting order parameter by $-2 \pi$ along a closed loop around one MBS with points $A, B$, and $C$ shown in (a). $\phi_{i=A,B,C}$ are the phases of the superconducting order parameter at positions $i$, given by $\phi_{A} = \phi_1, \phi_{B} = \phi_1-\pi +\delta \theta, \phi_{C} =  \phi_1-\pi -\delta \theta$, where $\delta \theta$ is a small positive number.}
	\label{sec4:fig2}       
\end{figure}

Figure~\ref{sec4:fig2}(a) shows a Corbino geometry JJ placed on the TI surface. The junction lying in the $x$-$y$ plane is made of thin films of inner ($S_1$) and outer ($S_2$) $s$-wave superconductors, which are connected by the TI surface. For simplicity, the distance between $S_1$ and $S_2$ is zero, which physically means that the distance is much smaller than the superconducting coherence length, forming a circular interface of radius $R$. If we introduce two flux quanta at the interface, the phase of the superconducting order parameter of $S_1$ remains constant, while the phase for $S_2$ enclosing the flux quanta winds by $-4 \pi$ \cite{Clem2010},  
\begin{equation}
\Delta(r, \theta)= 
\left\{ 
\begin{array}{cc}
\Delta_0 e^{i \phi_1} & 0 \leq r < R,\\
\Delta_0 e^{-i 2 \theta + i \phi_2} &  r > R,
\end{array}
\right. \label{sec4:winding}
\end{equation}  
where $\phi_1$ and $\phi_2$ are spatially uniform phases of the two superconductors. In a thin-film superconductor, there is a length $l$ which characterizes the spatial variation in phase across a JJ, and thus determines the relation between the phase associated with the winding number and $\theta$. Here, the phase factor $-i 2 \theta$ in equation~(\ref{sec4:winding}) is valid for $2 \pi R \ll l$. If one considers the case $2 \pi R \geq l$, the phase factor is no longer linear in $\theta$, although the winding by $-4\pi$ remains invariant \cite{Clem2010}, which would lead to quantitative modifications of subgap structures such as the localization width of a bound state and subgap level energies. However, the physics we discuss below results from the exchange statistics of MBSs and is not affected by such modifications.

The BdG Hamiltonian of the system has the form of Eq.~(\ref{sec4:BdGH}) with the proximity-induced gap described by equation~(\ref{sec4:winding}). 
The local superconducting phase difference across the junction is $\delta \phi(\theta) = \phi_1 - \phi_2 + 2 \theta$. Then, on the circumference of the junction, there are two positions $\theta_{\pm}$ satisfying $\delta \phi(\theta_{\pm}) = \pm \pi$, 
\begin{equation}\label{sec4:position}
\theta_{\pm} = \frac{\phi_2 - \phi_1 \pm \pi}{2}.
\end{equation}
As shown in Figure~\ref{sec4:fig2}(b), for each position, the order parameter along a closed loop containing the position has the winding number and thus a MBS is localized. Note that the phases $\phi_1$ and $\phi_2$ are gauge dependent, however, their difference is gauge invariant and determines the position of the MBSs. Since all positions along the circle are equally likely in terms of energy due to the rotation symmetry, only their relative position is fixed by the number of fluxes. Since $\phi_1$ and $\phi_2$ are the single-valued phases of the superconducting order parameters, their initial values, and thus the initial positions are chosen spontaneously.

The fact that the positions of MBSs are moved by changing $\phi_1 - \phi_2$ allows to perform an adiabatic exchange. It can be achieved experimentally by applying a dc voltage $V_J$ across the junction. Then $\phi_1 - \phi_2$ varies with time $t$ as $\phi_1 - \phi_2 = \phi_0 + 2 e V_J t/\hbar$, 
where $\phi_0$ is constant. During a half rotation time $T_J = \pi \hbar / (e V_J)$, the phase difference is changed by $2 \pi$ and the two MBSs are exchanged, 
\begin{equation}\label{sec4:exchange}
\gamma_1 \rightarrow -s \gamma_2, \,\,\,\,\, \gamma_2 \rightarrow s \gamma_1,
\end{equation}
governed by a braiding operator 
\begin{equation}
B = e^{\frac{\alpha}{2} \gamma_2 \gamma_1}. 
\end{equation} 
Here, $\alpha = s \pi/2$ is the exchange phase revealing the exchange statistics of MBSs. $s=1 (-1)$ corresponds to the change of $\phi_1 (\phi_2)$ by $2\pi (-2 \pi)$.

\subsection{Transport signatures of exchange statistics}
\label{sec4:measurement}

\begin{figure}
	\centering
	\resizebox{1\columnwidth}{!}{\includegraphics{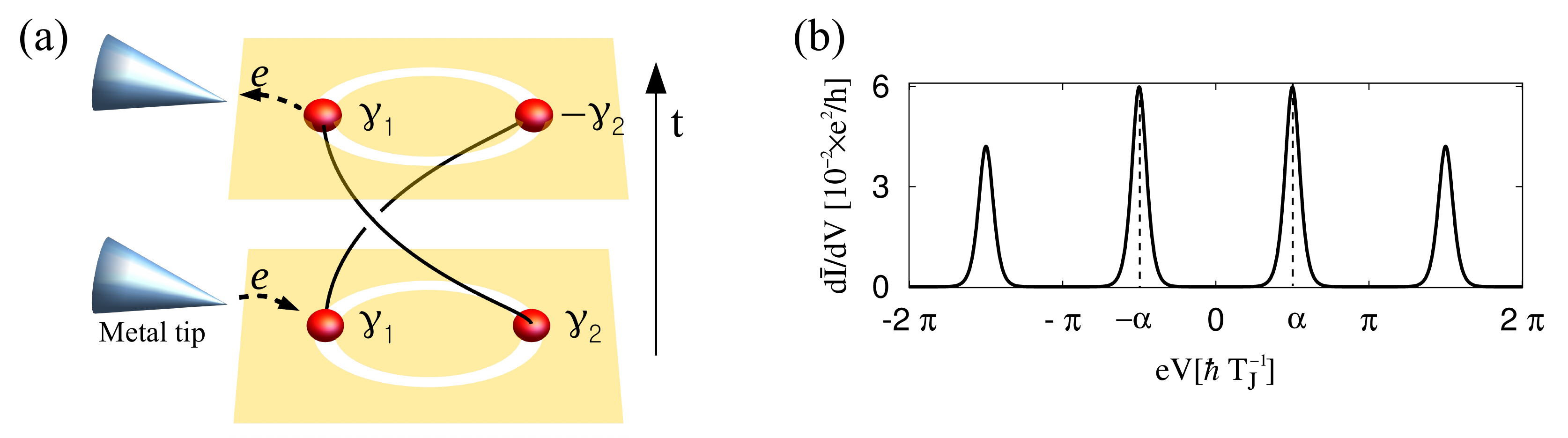}}
	\caption{(a) Time-dependent tunneling between a metal tip and two circulating MBSs (shown for a clockwise rotation). The tunneling occurs periodically with a period $T_J$, for which the two MBSs are exchanged. (b) Time-averaged differential conductance as a function of $eV$. The parameters are $\hbar T_{J}^{-1} = 0.1~\textrm{meV} = 10^{-1} \hbar \lambda = 10 k_B T = 10 \Gamma$. The exchange phase $\alpha =\pi/2$ is used. Around zero voltage, the peaks emerge at $eV = \pm \alpha \hbar T^{-1}_{J}$, constituting a direct imprint of the exchange phase of two MBSs. Figure (b) taken from~\cite{Park2015}.}
	\label{sec4:fig3}       
\end{figure}

We now couple a metal tip with bias voltage $eV$ to a position of the Corbino geometry JJ and analyze the tunneling conductance between the tip and MBSs travelling adiabatically along the junction to detect the exchange statistics. \footnote{The bias voltage can be defined relative to a grounded superconductor which can be either $S_1$ or $S_2$} Because the bound states decay exponentially away from the positions $(r,\theta)=(R,\theta_{\pm})$, one of them, which we set to $\gamma_1$, would couple to the tip. The coupling becomes significant and suppressed as $\gamma_1$ approaches to and leaves from the tip, respectively, leading to time-dependent tunneling events occurring periodically with period $T_J$. In each period, the MBSs are exchanged, see Figure~\ref{sec4:fig3}(a). Consequently, we define the discrete time sequence $t_q = t_0 + q T_J$ at which the coupling is maximal, where $q=0,1,2...$. The tunneling Hamiltonian in the low-energy subspace where only the zero-energy states contribute to the tunneling conductance is \cite{Park2015}
\begin{equation}
H_T(t) = \sum_{q,k} e^{-\lambda |t-t_q|} \left(V_{k} c^{\dagger}_{k} \gamma_1 +\textrm{H.c.}\right), \label{sec4:ht}
\end{equation}
where $\lambda^{-1}$ is the tunneling duration, $c^{\dagger}_{k}$ is the electron creation operator of the tip and $V_{k}$ is the coupling coefficient \footnote{It is sufficient to use spinless electrons $c^{\dagger}_{k}$ as the MBSs are spin-polarized and therefore couple only to one spin-direction of the tip electrons}

The resulting time-averaged tunneling current in the weak tunneling limit is plotted in Figure~\ref{sec4:fig3}(b), exhibiting peaks at distinct bias voltages, 
\begin{equation}
eV = \pm (\alpha - 2 \pi l) \hbar/T_J,
\end{equation}
where $l$ is an integer. This feature results from the coherent interference in the time-dependent tip-Majorana tunneling where the exchange phase $\alpha$ from the half-rotation of MBSs introduces a relative 
phase between the tunneling at time $t'$ and $t'+T_J$. The interference effect can also be understood by noting that tunneling of electrons from and to the tip changes the parity (e.g. $|0\rangle\rightarrow |1\rangle$), and therefore the braiding phase as outlined in the introduction. Suppose that $N$ half-rotations are interrupted by a tunneling event after $n<N$ half-rotations. This path can interfere with a path with $N$ half-rotations that is interrupted after $m<N$ half-rotations by a tunneling event (both paths change, e.g., $|0\rangle\rightarrow |1\rangle$). The probability for this event results from the absolute square of the superposition of these two amplitudes which have the same initial and final state leading to an exchange phase dependence $\propto \cos[(n-m)\alpha/2]$. We emphasize that, different to the case of tunneling between the tip and a static MBS which yields a conductance peak at zero bias voltage, we obtain conductance peaks at {\it non-zero voltage} for {\it zero-energy} MBSs due to the exchange operation. 
\section{Conclusion}
We gave a brief summary of recent efforts on the creation, detection and manipulation of MBSs in diverse quantum hybrid systems. We also briefly discussed their potential as topologically protected qubits owing to their non-Abelian braiding statistics. Subsequently, we then summarized four of our own projects regarding MBSs in detail.\\
First, we considered a spinless lead-MBS-quantum dot setup and showed that a pair of Fano resonances (FRs) arise in the differential conductance as a function of the quantum dot level energy $\varepsilon_D$ in specific parameter regimes. These FRs are symmetric under $\varepsilon_{D}\rightarrow-\varepsilon_{D}$ as long as lead and quantum dot only couple to one MBS, because of the particle-hole symmetry of a single MBS. As soon as we introduce a Dirac-like coupling to the quantum dot and the lead, i.e. a coupling to not only  the nearest but also to the more distant MBS, the symmetry of the FRs vanishes. Therefore, this symmetry can be used as a signature of coupling to a spatially isolated MBS.\\
Second, we summarized our findings for the equilibrium supercurrent in a JJ between an $s$-wave superconductor and a finite length Majorana nanowire. Due to the finite size of the Majorana nanowire not only the wave function of the closest MBS but also the wave function of the more distant MBS can have a finite weight at the junction. We analytically showed that the low energy contribution to the Josephson current is a function of the difference of the two spin canting angles of the MBS wave functions at the junction and is only finite if coupling to both MBSs is realized. This feature directly reflects the spin-singlet nature of the Cooper pairs mediating the Josephson effect. A numerical tight-binding model verifies the oscillatory behavior of the low energy contributions to the supercurrent as function of applied Zeeman field due to rotations of the spin canting angle of the more distant MBS.\\
Next, we considered a tunable JJ consisting of a sheet of silicene where the superconductivity is proximity induced by two superconducting leads. Two applied electric fields can be used to change the topology of the silicene sheet and thus tune the JJ between being topologically trivial and topologically non-trivial. In the topologically non-trivial regime the Josephson current is mediated by spin-helical edge states and is 4$\pi$ periodic as a function of the phase difference between the two superconducting leads due to the formation of MBSs.\\
Finally, we analyzed a setup consisting of a Corbino geometry JJ on the surface of a 3D TI. The introduction of two magnetic flux quanta into the junction results in the formation of two MBSs which can rotate along the junction when a voltage is applied across the junction due to the temporally changing superconducting phase difference. We showed that the time-averaged differential conductance measured using a metallic tip positioned at the junction reveals the non-trivial MBS statistics.\\
Open access funding provided by Projekt DEAL. We gratefully acknowledge the support of the Lower Saxony PhD-programme Contacts in Nanosystems, the Braunschweig International Graduate School of Metrology B-IGSM and the DFG Research Training Group 1952 Metrology for Complex Nanosystems, the Hannover-Braunschweig science cooperation QUANOMET and DFG-EXC 2123, Quantum Frontiers.
\section*{Author contribution statement}
AS performed the calculations presented in Sections~\ref{Sec:Fano} and~\ref{Sec:spincanting}. DF executed the computations shown in Section~\ref{Sec:silicene}. SP was in charge of Section~\ref{sec4:Corbino}.  All authors wrote the initial manuscript of the sections for which they performed the calculations. PR wrote the manuscript for Section~\ref{intro} and parts of Section~\ref{sec4:Corbino}. PR supervised all projects. All authors provided feedback and helped to shape the manuscript.\\
\textbf{Open Access} This is an open access article distributed under the terms of the Creative Com-
mons Attribution License (\href{http://creativecommons.org/licenses/by/4.0}{http://creativecommons.org/licenses/by/4.0}), which permits unrestricted
use, distribution, and reproduction in any medium, provided the original work is properly cited.

\end{document}